\documentclass[%
 reprint,
superscriptaddress,
amsmath,amssymb,
aps,
]{revtex4-2}

\usepackage{graphicx}% Include figure files
\usepackage{dcolumn}% Align table columns on decimal point
\usepackage{bm}% bold math
\usepackage{microtype}
\usepackage{url}
\usepackage{xcolor}
\usepackage{color}
\definecolor{amaranth}{rgb}{0.9, 0.17, 0.31}
\definecolor{palatinateblue}{rgb}{0.15, 0.23, 0.89}
\definecolor{radicalred}{rgb}{1.0, 0.21, 0.37}
\usepackage{hyperref}
\usepackage[utf8]{inputenc}
\usepackage[T1]{fontenc}
\hypersetup{ linktoc=all,
	colorlinks, linkcolor={palatinateblue},
	citecolor={radicalred}, urlcolor={amaranth}
}

\usepackage{mathrsfs}

\usepackage{cleveref}

\renewcommand{\eqref}[2][]{Eq{#1}.~(\ref{eq:#2})}		% Equation reference
\newcommand{\lb}{\ensuremath{\left}}					% Left Brackets
\newcommand{\rb}{\ensuremath{\right}}					% Right Brackets
\newcommand{\gpgg}{\ensuremath g_{\phi\gamma}}

\newcommand{\nhat}[1][]{\bm{\hat{n}^{#1}}}

\def\jcap{J.\ Cosmol.\ Astropart.\ Phys.\ }
  \newcommand{\beq}{\begin{equation}}
  \newcommand{\eeq}{\end{equation}}
  
  \newcommand{\bi}{\begin{itemize}}
  \newcommand{\ei}{\end{itemize}}
\begin{document}

\title{Constraining Ultralight ALP Dark Matter in Light of Cosmic Birefringence
}

\author{Dongdong Zhang}
\email[Email: ]{don@mail.ustc.edu.cn}
\affiliation{Department of Astronomy, School of Physical Sciences, University of Science and Technology of China, Hefei, Anhui, 230026, China}
\affiliation{School of Astronomy and Space Science, University of Science and Technology of China, Hefei, Anhui, 230026, China}
\affiliation{CAS Key Laboratory for Researches in Galaxies and Cosmology, University of Science and Technology of China, Hefei, Anhui, 230026, China}
\affiliation{Kavli Institute for the Physics and Mathematics of the Universe (WPI), UTIAS, The University of Tokyo, Chiba 277-8583, Japan}

\author{Elisa G. M. Ferreira}
\email[Email: ]{elisa.ferreira@ipmu.jp}
\affiliation{Kavli Institute for the Physics and Mathematics of the Universe (WPI), UTIAS, The University of Tokyo, Chiba 277-8583, Japan}

\author{Ippei Obata}
\email[Email: ]{ippei.obata@ipmu.jp}
\affiliation{Kavli Institute for the Physics and Mathematics of the Universe (WPI), UTIAS, The University of Tokyo, Chiba 277-8583, Japan}

\author{Toshiya Namikawa}
\email[Email: ]{toshiya.namikawa9@ipmu.jp}
\affiliation{Center for Data-Driven Discovery, Kavli Institute for the Physics and Mathematics of the Universe (WPI), UTIAS, The University of Tokyo, Chiba 277-8583, Japan}

\begin{abstract}
Cosmic birefringence, the observed rotation of the polarization plane of the cosmic microwave background (CMB), serves as a compelling probe for parity-violating physics beyond the Standard Model.
This study explores the potential of ultralight axion-like particle (ALP) dark matter to explain the observed cosmic birefringence in the CMB.
We focus on the previously understudied mass range of $10^{-25}$ eV to $10^{-23}$ eV, where ALPs start to undergo nonlinear clustering in the late universe.
Our analysis incorporates recent cosmological constraints and considers the washout effect on CMB polarization.
We find that for models with ALP masses $10^{-25}$ eV $\lesssim m_\phi \lesssim 10^{-23}$ eV and birefringence arising from late ALP clustering, the upper limit on the ALP-photon coupling constant, imposed by the washout effect, is stringently lower than the coupling required to account for the observed static cosmic birefringence signal.
This discrepancy persists regardless of the ALP fraction in dark matter.
Furthermore, considering ALPs with masses $m_\phi\gtrsim$ $10^{-23}$ eV cannot explain static birefringence due to their rapid field oscillations, our results indicate that, all ALP dark matter candidates capable of nonlinear clustering in the late universe and thus contributing mainly to the rotation angle of polarized photons, are incompatible with explaining the static cosmic birefringence signal observed in Planck and WMAP data.

\end{abstract}

\maketitle
\section{\label{Sec1}Introduction}
Cosmic birefringence, the rotation of the polarization plane of the CMB as it travels through space, has emerged as a new observational effect in cosmology. This phenomenon offers a unique avenue to probe parity-violating physics in the universe. Recent analyses of cross-correlations between the even-parity CMB E-modes and odd-parity CMB B-modes, have provided intriguing evidence for isotropic cosmic birefringence with a rotation angle $\beta \sim 0.35$ deg in $3.6\sigma$ C.L. \citep{2020PhRvL.125v1301M,2022PhRvL.128i1302D,2022A&A...662A..10E,2022PhRvD.106f3503E,2023A&A...679A.144E,2022NatRP...4..452K}. The {\it Planck}\ space observatory and ground-based experiments like \textsc{BICEP}, \textsc{ACTPol} and \textsc{SPTpol} have placed stringent constraints on anisotropic cosmic birefringence \citep{BICEP2:2017lpa,BK-LoS:2023,2020PhRvD.101h3527N,2020PhRvD.102h3504B,Bortolami:2022whx,Zagatti:2024,Namikawa:2024:BB}.
Additionally, the measurement of the Crab Nebula by \textsc{POLARBEAR} has
suggested a nonzero birefringence signal \citep{2024arXiv240302096A}. 

One promising explanation for isotropic cosmic birefringence involves pseudoscalar fields such as ALPs \citep{2020arXiv200802473F}.
These particles interact with electromagnetic fields through a Chern-Simons coupling, leading to a rotation of the polarization plane of light \citep{Harari:1992ea,1998PhRvL..81.3067C}. 
The ALP model that induces birefringence has been extensively studied in various cosmological contexts, including dark energy (DE) \citep{1998PhRvL..81.3067C,2011PhRvD..83h3506P,2020arXiv200802473F,2021PhRvD.103d3509F,2021PhRvD.104j1302C,2022JCAP...09..062O,2022JCAP...08..025G}, early dark energy \citep{2021PhRvD.103d3509F,2023PhRvD.107d1302M,2023PhRvL.131l1001E}, cosmic domain-wall networks \cite{Takahashi:2020tqv,Kitajima:2022jzz,Gonzalez:2022mcx,Kitajima:2023kzu,Ferreira:2023jbu} and dark matter (DM) scenarios \citep{2009PhRvD..79f3002F,2017PDU....16...22L,2019PhRvD.100a5040F,Murai:2024yul}.
Each context introduces different dynamics and observational signatures for the ALP field.
Understanding the behavior of ALPs across different mass ranges is crucial for interpreting the cosmic birefringence signal.
Multiple tomographic approaches with different observables may facilitate this understanding~\cite{Sherwin:2021vgb,Nakatsuka:2022epj,Lee:2022udm,2023PhRvD.107h3529G,Naokawa:2023upt,Namikawa:2023zux,Naokawa:2024xhn}.
By analyzing the CMB and other cosmological data, we can constrain the properties of ALPs and refine our models of cosmic birefringence.

In this work, we calculate the capability of ALPs to account for cosmic birefringence in the long-neglected mass range of $10^{-25}$ eV to $10^{-23}$ eV and point out that this possibility is stringently excluded by the washout effect in the CMB polarization. We find that the ALP mass ranges $m_\phi \gtrsim 10^{-25}$~eV, which is widely considered capable of producing late-time non-linear structure, fails to explain the isotropic birefringence. Therefore, if ALP dark matter (ADM) candidates are introduced, and if the cosmic birefringence signal is further confirmed in future experiment releases, it would imply the necessity of non-conventional potentials, multiple ALP models or more complex theoretical frameworks.

The structure of this paper is as follows.
In Section \ref{Sec2}, we briefly introduce cosmic birefringence and ALP models and calculate the constraints on the non-linear evolution ADM model from cosmic birefringence.
In Section \ref{Sec3}, we present the current cosmological and astronomical constraints on the axion fraction within our mass range of interest, calculate the constraints on the ALP coupling constant from the CMB washout effect in this mass range, and compare them with the cosmic birefringence constraints.
The conclusions are presented in Section \ref{Sec4}.

\section{\label{Sec2}Cosmic Birefringence and ALP model}

\subsection{\label{Subsec2.1}Birefringence by ALPs}

The Standard Model of particle physics (SM), despite its remarkable success in describing the fundamental interactions of nature, is known to be incomplete. Many theories beyond SM predict the existence of new fields and particles that could have observable effects on the Universe. One such effect is cosmic birefringence, and previous research has shown that this phenomenon must be explained beyond the framework of SM \citep{2024JHEP...01..057N}.

Axions were originally proposed in the 1970s by  Peccei and  Quinn \citep{Peccei:1977hh,Peccei:1977ur}, Weinberg \cite{Weinberg:1977ma} and Wilczek \cite{Wilczek:1977pj}, as a solution to the ``strong CP problem" in quantum chromodynamics (QCD), called ``QCD axion". This problem addresses the absence of observed CP violation in the strong interactions of fundamental particles.
QCD axion emerges from the spontaneous breaking of the Peccei-Quinn symmetry, leading to their hypothesized role as a component of dark matter. They are characterized by very low masses and weakly interacting capabilities.

ALPs are hypothesized as a class of pseudoscalar particles, akin to the QCD axion, yet distinguishable by their less constrained mass and coupling parameters, allowing them to evade some of the stringent experimental bounds that apply to conventional axion models.
Theoretically, due to the parity-breaking interaction between ALPs and photons, linearly polarized photons can undergo polarization plane rotation when passing through an ALP field, exhibiting birefringence.
The CMB photons are sensitive probes of this effect. 
The rotation angle $\beta$, which is imprinted on CMB photons emitted from the last scattering surface (LSS) and subsequently reaching an observer, is sourced by the ALP field displacement at the LSS and the observer's position. This angle is directly proportional to the coupling constant $g_{\phi\gamma}$:
\begin{equation}
\beta(\hat{\bm{n}}) = \frac{g_{\phi\gamma}}{2} \left[ \phi(t_0, \bm{0}) - \phi(t_{\mathrm{LSS}}, d_{\mathrm{LSS}}\hat{\bm{n}}) \right],
\end{equation}
where $\bm{0}$ denotes the observer's position, $t_0$ signifies the present time, $t_{\mathrm{LSS}}$ the epoch of last scattering, and $d_{\mathrm{LSS}}$ the distance to the LSS. The ALP field $\phi$ encompasses both the background and fluctuated components, indexed ``0" for the current time and ``LSS" for the time at last scattering:
\begin{align}
    \phi (t_{\mathrm{LSS}}, d_{\mathrm{LSS}}\hat{\bm{n}}) &= \bar{\phi}_{\mathrm{LSS}} + \delta\phi_{\mathrm{LSS}}, \label{eq:AnisotropicBirefringence}\\
    \phi (t_0, \bm{0}) &= \bar{\phi}_{\mathrm{obs}} + \delta\phi_{\mathrm{obs}}.
\label{eq:IsotropicBirefringence}
\end{align}
Consequently, the rotation angle $\beta$ is dissected into isotropic and anisotropic components:
\begin{multline}
    \beta(\hat{\bm{n}})
    =
    \bar{\beta} + \delta \beta(\hat{\bm n}) \\
    =
    \frac{g_{\phi\gamma}}{2} \left(\bar{\phi}_{\mathrm{obs}} - \bar{\phi}_{\mathrm{LSS}} + \delta \phi_{\mathrm{obs}}\right) -     \frac{g_{\phi\gamma}}{2} \delta \phi_{\mathrm{LSS}}(\hat{\bm n}).
    \label{alpha full expression}
\end{multline}

\subsection{\label{Subsec2.2}ALPs as Dark Matter}

When ALPs are nearly frozen due to Hubble friction, their equation of state, $w_\phi$, approaches $-1$.
Once ALPs begin to oscillate, they behave like dust with an average equation of state $w_\phi \sim 0$, making them potential candidates for DM.
ALPs capable of acting as DM span a wide mass range, from $10^{-32} \, \mathrm{eV}$ to $\mathrm{eV}$.
When $m_{\phi} > 10^{-28} \,\mathrm{eV}$, and the mass term approximates $H(t_{\mathrm{eq}})$, ALPs exhibit coherent oscillations after matter-radiation equality. They behave like non-relativistic matter and can serve as DM candidates.
For masses smaller than $10^{-28}$ eV, ALPs contribute to DE for a period after equality.
For $m_\phi \lesssim 10^{-34}\, \mathrm{eV}$,  where the upper limit is determined by the current Hubble constant ($m_\phi \sim 3 H_0 \sim 10^{-33}$ eV) and the constraint on the equation of state $w_\phi$ from {\it Planck} data \citep{2020A&A...641A...6P,2021PhRvD.103d3509F}, ALPs are allowed to form a substantial fraction of DE.
Previous studies on cosmic birefringence have primarily focused on ALPs with $m_{\phi} \lesssim 10^{-27}$ eV. In this mass range, ALPs behave like DE or DM. For the ADM in this mass range, the de Broglie wavelength significantly exceeds galactic scales, rendering its nonlinear gravitational clustering effects negligible. Therefore, apart from contributing to isotropic birefringence, the only contribution to anisotropic birefringence comes from linear perturbations present at recombination.

However, in the mass range $m_{\phi} \gtrsim 10^{-25}$ eV, ALPs exhibit clustering in the late universe, participating in the formation of the non-linear structures in our universe, including the dark matter halo. Therefore, the local dark matter density is significantly higher than the cosmic average. 
This is confirmed by ALPs simulations \citep{2024PhRvD.109d3507L,Schwabe:2020eac} indicating that the density of ALP particles in our location in the Milky Way (MW), compared to the average density of ALPs in the current universe, does not differ significantly from that of CDM. 
Thus, a local non-zero ALP field in this mass range could potentially explain the static isotropic cosmic birefringence.
ADM with a mass smaller than this also clusters and participates in the formation of the LSS of the universe. Given their large de Broglie wavelength and the fact that cosmological constraints only allow this to be a small part of DM~\cite{Hlozek:2014lca,Hlozek:2017zzf,Lague:2021frh,Rogers:2023ezo}, ADM has a less important role in the formation of the halo.

Reference \citep{2019arXiv190513650C} pointed out that when considering the local ALP field within the MW as the sum of many individual wave modes, each with random phase, the amplitude $\phi_{\mathrm{loc}}$ at any given time is a Rayleigh-distributed stochastic variable centered on $\phi_0$ that varies with time on the ALP coherence timescale $t_{\text{coh}} \sim 2\pi/(m_\phi v_0^2)$, where $v_0$ is the MW virial velocity, as also discussed in \citep{Nakatsuka:2022gaf,2023PhRvD.108d3017A}. 
We fix $\phi_0$ such that the local ALP density $\rho_{0\mathrm{a}}$ is a fraction $\kappa$ of the local dark matter density $\rho_{\mathrm{0dm}}$. With $\rho_{\mathrm{0dm}} \approx 0.4 ~ \text{GeV}/\text{cm}^3$ \citep{2015NatPh..11..245I,2018MNRAS.478.1677S} and $\kappa \equiv \rho_{0\mathrm{a}} / \rho_{\mathrm{0dm}}$, we have:

\begin{equation}
\frac{1}{2} m_\phi^2 \phi_0^2 = \kappa \rho_{0\mathrm{dm}}.
\label{eq:phi0}
\end{equation}
Therefore, the center ALP field amplitude is:
%%%
\begin{align}
        \phi_0 &= \sqrt{\frac{2\rho_{0\mathrm{dm}}
        \kappa}{m_{\phi}^2}} \\
        &\approx2.5\times 10^{12}~\text{GeV} \,
       \kappa^{1/2}
        \left(\frac{m_{\phi}}{10^{-24}\,\text{eV} } \right)^{-1}\,.
    \label{eq_localALPphi}
\end{align}
%%%
The possibility function of the Rayleigh distribution $P(\phi)$ is:
\begin{equation} \label{eq:Pphi}
    P(\phi_{\mathrm{loc}}) = \frac{2 \phi_{\mathrm{loc}}}{\phi_0^2} e^{-\frac{\phi_{\mathrm{loc}}^2}{\phi_0^2}}.
\end{equation}
Thus, the local ALP field contributes to isotropic birefringence via the $\delta\phi_\mathrm{obs}$ term in Eq.~(\ref{eq:IsotropicBirefringence}). Using the 95\% upper limit of the field amplitude, $\phi_{\mathrm{loc}}=1.73 \, \phi_0$, and assuming that $\bar{\phi}_{\mathrm{LSS}}$ in Eq.~(\ref{eq:IsotropicBirefringence}) is canceled due to the fast oscillating ALP field, we estimate the coupling constant needed to explain cosmic birefringence. This coupling constant can be determined as:

\begin{multline}
    g_{\phi\gamma} = \frac{2\bar\beta}{\phi_{\mathrm{loc}}} \\
        \approx
        2.8 \times 10^{-15}\,\text{GeV}^{-1}
        \left(\frac{\bar\beta }{0.35\,\text{deg}} \right)
        \kappa^{-1/2}
        \left(\frac{m_{\phi}}{10^{-24}\,\text{eV} } \right).
    \label{eq:localbirefringence}
\end{multline}
Eq.~(\ref{eq:localbirefringence}) indicates that the ALP-photon coupling constant required to explain the birefringence angle is significantly lower than the constraints set by CAST~\citep{2017NatPh..13..584A}.
Additionally, the effects of washout are also worth considering.

\section{\label{Sec3}Data Constraints}

\subsection{\label{Subsec3.1} Current constraints on ADM}

Recent advancements in cosmological and astrophysical observations have provided significant constraints on the potential role of ultralight ALPs as dark matter across various mass ranges. This subsection summarizes the latest data limitations and the scope of ALP dark matter research. A review of the constraints on this model can be seen in the reviews~\cite{NIEMEYER2020103787,2021A&ARv..29....7F, Hui:2021tkt,Adams:2022pbo} or in {\tt \href{https://keirkwame.github.io/DM_limits/}{this compilation}}.

\begin{figure}
    \centering
    \includegraphics[width=1.05\linewidth]{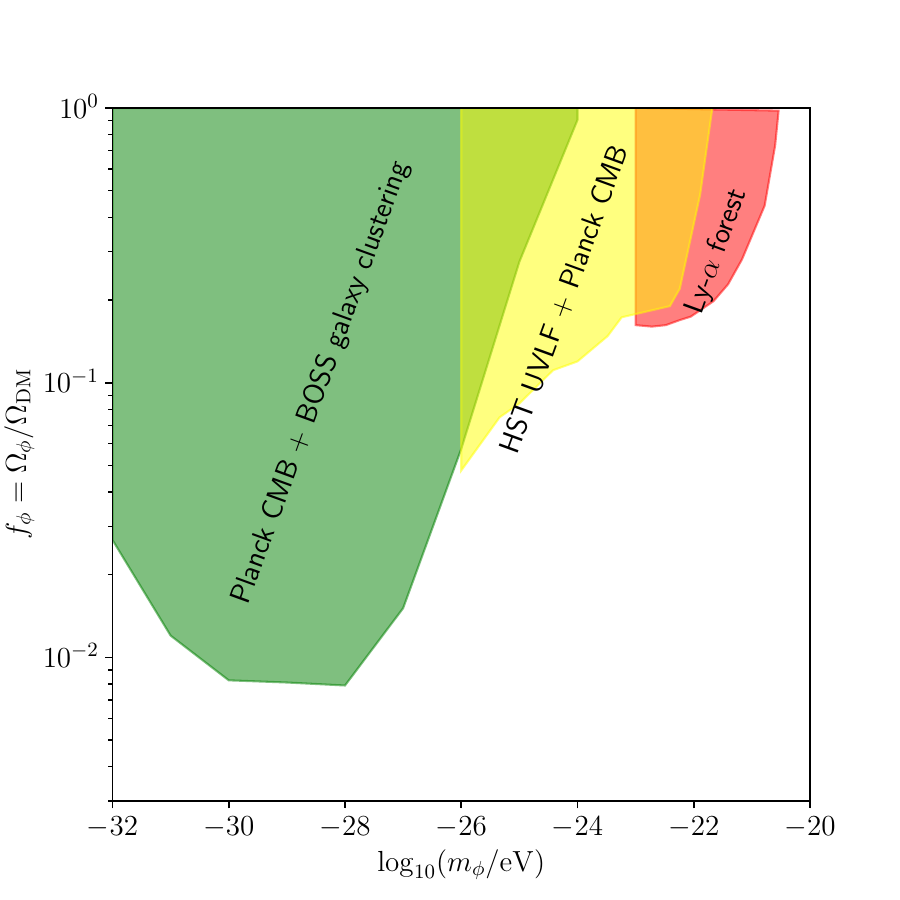}
        \caption{
    Bounds in the mass and fraction of ADM in the mass range of interest for this work. Figure adapted from~\cite{2024arXiv240411071W}.
}
\label{fig:f_constraints}
\end{figure}

The most important and stringent constraints we have in ALPs come from the CMB and Large Scale Structure showing that $m_\phi > 10^{-24}\,\mathrm{eV}$ for ALPs to be all the DM~\cite{Hlozek:2014lca,Hlozek:2017zzf,Lague:2021frh,Rogers:2023ezo}. For masses smaller than that, ADM can only represent a fraction ($f_\phi = \Omega_\phi / \Omega_{\mathrm{DM}}$, where $\Omega_x= \rho/\rho_c$ is the density parameter, with $\rho_c = 3H^2/8\pi G$ the critical density) of the total dark sector, with this fraction changing depending on the mass range. Another strong constraint in the ALP mass and fraction comes from the Lyman-$\alpha$ forest  \cite{Irsic:2017yje,Kobayashi:2017jcf,Rogers:2020ltq} where $m_\phi \gtrsim 10^{-20} \, \mathrm{eV}$ if ALPs are $100\%$ of the dark matter (with differences between the values of this bound depending on the reference). For smaller masses, ALPs can only be a fraction of DM and with data from XQ-100 survey and the MIKE and HIRES spectrographs~\cite{2016A&A...594A..91L}, one can bound the fraction of ALP allowed from $ 10^{-23} \, \mathrm{eV} \lesssim m \lesssim 10^{-20} \, \mathrm{eV}$~\cite{2024arXiv240411071W}. Recently, using 
the galaxy UV luminosity function (UVLF) as an independent probe of ultralight ALPs obtained from high-redshift galaxy observations from the Hubble and James Webb Space Telescopes, the fraction of ALPs in the mass range $10^{-26}~\text{eV} \lesssim m_\phi \lesssim 10^{-23}~\text{eV}$ was constrained~\cite{2024arXiv240411071W}. In Fig.~(\ref{fig:f_constraints}) we summarize the bounds cited above since they are the relevant ones for this work.

ALPs with masses in the $10^{-22} \quad \mathrm{eV} \lesssim m_{\phi} \lesssim 10^{-17} \quad \mathrm{eV}$ range fall within the mass range of so-called ``fuzzy dark matter" (FDM). In this mass range, FDM can be most if not all the DM in the universe, which is what we consider for most of the bounds we are discussing next. 
This model has been extensively studied with important observational constraints \cite{NIEMEYER2020103787,2021A&ARv..29....7F,Hui:2021tkt} that have narrowed down its mass range.
Besides the CMB+Large Scale Structure and the Lyman-$\alpha$ forest constraints, here are some of the most important or stringent bounds in this model. Using stellar kinematic data from Milky-Way dwarf galaxies we can constraint the DM density profile in these systems, showing a preference for larger FDM masses~\cite{Hayashi:2021xxu,Zimmermann:2024xvd,Chan:2021bja}.
There are also constraints targeting the interference patterns produced by these models. Strong lensing, is a gravitational probe sensitive to the presence of granules, that show that $m_\phi \gtrsim 10^{-21} \, \mathrm{eV}$ from different analysis~\cite{Powell:2023jns,Laroche:2022pjm}, while a different bound was found in \cite{Amruth:2023xqj} with a preference for lower masses.
The presence of the granules also causes stellar heating and its constraint on the stellar kinematic data from ultrafaint dwarf galaxies leads to a very strong bound on the mass of FDM, $m_\phi > 3 \times 10^{-19} \, \mathrm{eV}$~\cite{Dalal:2022rmp}.  We cite here only some of the strongest ones, with many others present in the literature. Some of these also constrain the coupling constant like using the lack of CMB polarization suppression \citep{2019PhRvD.100a5040F} and the effects of oscillations in the local ALP field \citep{2023PhRvD.108d3017A,2022PhRvD.106d2011F,2022PhRvD.105b2006A}. Because the oscillation frequency of the local ALP field is already large enough in this mass range, it cannot serve as a possible explanation mechanism for static cosmic birefringence.

\subsection{\label{Subsec3.2}Washout Effect}

In the context of ADM, the washout effect refers to a reduction in the polarization of CMB light compared to standard predictions.
During the CMB decoupling epoch, the ALP field with a mass range of $m_{\phi} \gtrsim 10^{-26}$ eV oscillates multiple times.
These oscillations lead to the ``washout" of the polarization signal because the polarization angle rotates back and forth, resulting in an averaging effect that reduces the observed net polarization.
For ALP dark matter with faster oscillation periods, the washout effect becomes more significant.
Fedderke et al. \citep{2019PhRvD.100a5040F} calculated the washout effect and the impact of local ALP field oscillations on the CMB polarization pattern under the FRW metric, assuming a finite thickness of the LSS:
%%%
\begin{widetext}
\begin{align}
&(Q \pm i U)(\hat{\bm{n}})\nonumber \\
&\approx \epsilon I \int d\tilde{t}' g(\tilde{t}') \exp\lb[ \pm 2i \lb(\psi +  \frac{\gpgg}{2} \phi_0 \cos(m_\phi t + \alpha ) - \frac{\gpgg}{2} \phi_*(\nhat)  \lb[ \frac{1+\tilde{z}}{1+z_*} \rb]^{3/2} \cos( m_\phi \tilde{t}' + \beta(t,\tilde{t}') ) \rb) \rb] \label{eq:ToyModel3Result1}\\
&\approx \epsilon I  \sum_n g_n \int_{\tilde{t}'_n-\delta t'/2}^{\tilde{t}'_n + \delta t'/2} d\tilde{t}' \exp\lb[ \pm 2i \lb(\psi +  \frac{\gpgg}{2} \phi_0  \cos(m_\phi t + \alpha ) - \frac{\gpgg}{2} \phi_*(\nhat) \lb[ \frac{1+\tilde{z}_n}{1+z_*} \rb]^{3/2}  \cos( m_\phi \tilde{t}' + \beta_n ) \rb) \rb] \label{eq:ToyModel3Result2}\\
&= J_0 \lb[ \gpgg \langle\phi_*\rangle(\nhat) \rb] \exp\lb[ \pm 2i \lb( \frac{\gpgg}{2} \phi_0 \cos(m_\phi t + \alpha ) \rb) \rb] (Q\pm i U)_0(\hat{\bm{n}}), 
\label{eq:ToyModel3Result3}
\end{align}
\end{widetext}
%%%%%%
where $(Q\pm i U)_0$ is the value measured in the ALP decoupling limit $\gpgg \rightarrow 0$, $\epsilon$ represents the intrinsic polarization asymmetry of the source, $g(\tilde{t}')\equiv g(\tilde{z}) ( d\tilde{z} / d\tilde{t}' )$ for convenience, and the average value of the ALP field amplitude weighted by the visibility function at decoupling:
%%%%%%
\begin{align}
&J_0 \lb[ \gpgg\langle \phi_* \rangle(\nhat)  \rb] \nonumber \\
&\equiv \int d\tilde{z}\, g(\tilde{z})\, J_0 \lb( \gpgg \phi_*(\nhat) \lb[ \frac{1+\tilde{z}}{1+z_*} \rb]^{3/2} \rb)\label{eq:AverageDefn1} \\
&\approx J_0 \lb[ \gpgg \phi_*(\nhat)  \rb].
\label{eq:AverageDefn2}
\end{align}
%%%%%%
The latter approximate equation holds because $g(\tilde{z})$ has a strong peak near $\tilde{z}=z_*$.
We can see that the amplitude of the local oscillation effect is clearly controlled by the amplitude of the local ALP field value, while the amplitude of the washout effect is controlled by the visibility function weighted average value of the ALP field amplitude at last scattering.
At the 95\% confidence level, Fedderke et al. \citep{2019PhRvD.100a5040F} concluded the constraints from the Planck 2018 data as follows:
\begin{equation}
g_{\phi\gamma}\lesssim 9.6 \times 10^{-16} \text{GeV}^{-1} 
\times \frac{m_{\phi}}{10^{-24} \text{eV}}
\times f_{\mathrm{\phi}}^{-1/2}.
\label{eq:washout}
\end{equation}

In our universe, LSS corresponds to a redshift of approximately $z \sim 1000$ to $1200$.
This equates to a cosmic time interval of approximately $10^5$ years, which is significantly longer than the oscillation period of ALPs within the mass range of interest.
Since our research extends the ALP mass range to lower values than those considered in \cite{2019PhRvD.100a5040F}, we must evaluate the validity of the approximations used in Eq.~(\ref{eq:ToyModel3Result2}).

\begin{figure}
    \centering
    \includegraphics[width=1.0\linewidth]{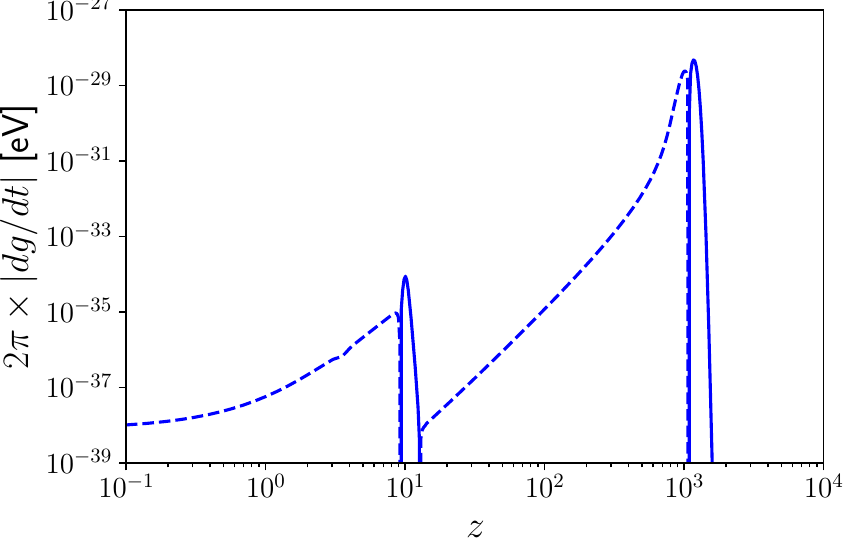}
        \caption{
    The partial derivative of the visibility function with respect to time. The ordinate is $2\pi (1+z)H(z)|\partial_z g(z)|=2\pi |\partial_t g(t)|$, and the abscissa is the redshift. The dashed line means a negative value. The curve in the figure was obtained using {\tt \href{https://camb.readthedocs.io/en/latest/}{pycamb}} \citep{Lewis:1999bs}, with cosmological parameters based on the best fit from Planck 2018 \citep{2020A&A...641A...6P}.
}\label{fig:visibility}

\end{figure}

When $2\pi |\partial_t g(t)| \ll m_\phi$ is satisfied, $g(t)$ remains fairly constant over one ALP period.
Fig.~(\ref{fig:visibility}) shows that if $m_{\phi} \gtrsim 10^{-27}$ eV, $g(t)$ can always be considered constant within an ALP cycle.
Additionally, within a time interval $\delta\tilde{t} = 2\pi/m_\phi$ near $z=z_*$, where $10^{-25}$ eV $\lesssim m_\phi \lesssim 10^{-23}$ eV and $\tilde{z} \equiv z(\tilde{t}')$, the condition $\delta \tilde{z} \equiv z(\tilde{t}'+\delta\tilde{t}') - z(\tilde{t}') \ll \tilde{z}$ remains valid. 
Therefore, in Eq.~(\ref{eq:ToyModel3Result1}), we can still write $1+\tilde{z} \approx 1+\tilde{z}_n$ in each subdomain.
Consequently, we conclude that in the mass range $10^{-25}$ eV $\lesssim m_\phi \lesssim 10^{-23}$ eV, Eq.~(\ref{eq:washout}), derived from approximate calculations, remains valid.

While this paper was being finalized, the following paper appeared~\citep{Murai:2024yul} where the validity of the analytical washout formula Eq.~(\ref{eq:ToyModel3Result3}) is confirmed for the ALP mass range $m_{\phi} > 10^{-26}\, \mathrm{eV}$ using the Boltzmann equation. 
Their work also discussed a non-conventional asymmetric ALP potential, which allows $\bar{\phi}_{\mathrm{LSS}}$ in Eq.~(\ref{eq:IsotropicBirefringence}) to retain sufficient values under rapid oscillations of ADM, and can still explain static cosmic birefringence considering the constraints of washout effect.
In this paper, we only consider the cases of the conventional ALP potentials.

\begin{figure}
    \centering
    \includegraphics[width=1.0\linewidth]{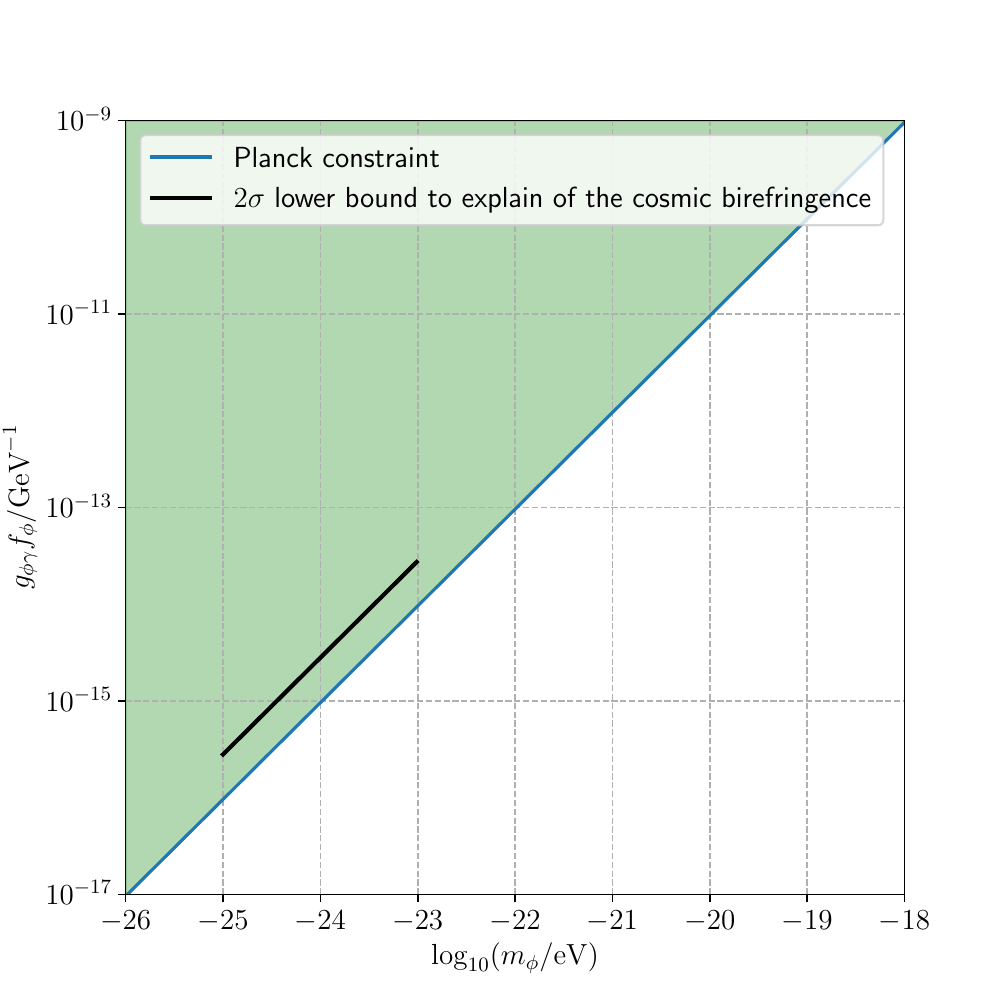}
        \caption{
    Comparison of the constraints on ADM from the washout effect after expanding the axion mass range compared to \citep{2019PhRvD.100a5040F}, with the coupling constant - mass relationship required to explain static cosmic birefringence at the 95\% C.L. lower bound. The abscissa represents the ALP mass, while the ordinate shows the coupling constant multiplied by the square root of the ALP fraction.
}\label{fig:coupling}
\end{figure}

\begin{figure}
    \centering
    \includegraphics[width=1.1\linewidth]{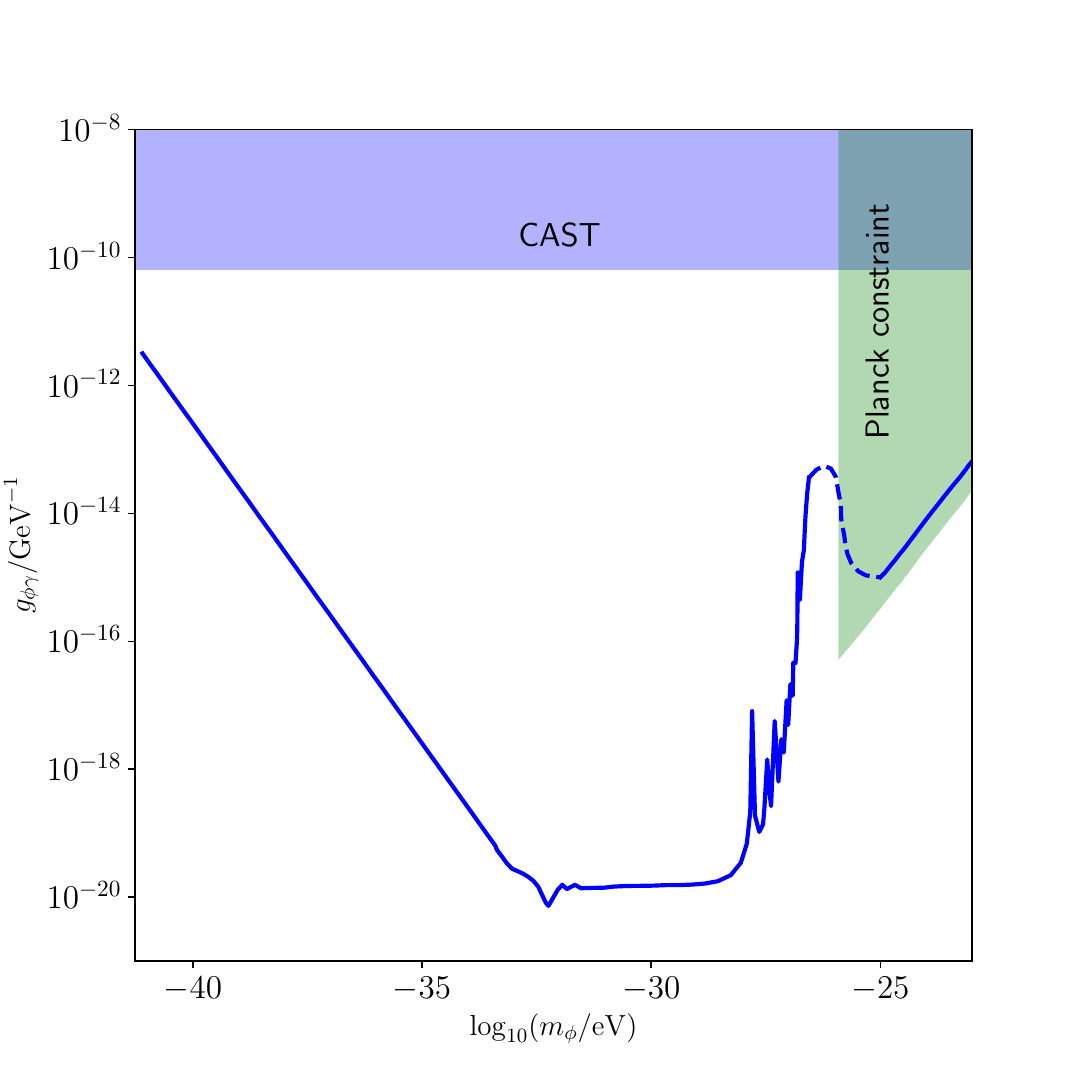}
        \caption{ALP-photon coupling versus ALP mass obtained from isotropic cosmic birefringence $\beta \sim 0.35 \deg$. The curve for $m_\phi \lesssim 10^{-26}$ eV is obtained from previous study \citep{2021PhRvD.103d3509F}, assuming ALP potential function $V(\phi) = m_\phi^2\phi^2/2$.
}\label{fig:fullcons}
\end{figure}

\subsection{\label{Subsec3.3}Analysis and discussion}

When the ALP mass $m_{\phi} \gtrsim 10^{-25}$~eV, the non-linear clustering behavior of ADM in our position in the MW, in our concerned mass ranges, does not significantly differ that of cold dark matter \citep{2024PhRvD.109d3507L,Schwabe:2020eac}, so it can be assumed that $\kappa=f_{\mathrm{\phi}}$, and the value of $\kappa$ in different mass ranges also follows Fig.~(\ref{fig:f_constraints}).
As discussed in the previous subsection, Eq.~(\ref{eq:ToyModel3Result3}) is valid for $m_{\phi} \gtrsim 10^{-26}$~eV, so according to Eq.~(\ref{eq:washout}), we can obtain the limit curve as shown in Fig.~(\ref{fig:coupling}), where the value of $f_{\mathrm{\phi}}$ is based on the constraint on the axion's fraction of dark matter in Fig.~(\ref{fig:f_constraints}).
In Fig.~(\ref{fig:coupling}), we also present the relationship between the coupling constant and mass required to explain the isotropic birefringence signal according to Eq.~(\ref{eq:localbirefringence}).

We constrain the mass range to $m_{\phi} < 10^{-23}$~eV, because for $m_{\phi} \gtrsim 10^{-23}$~eV, the oscillation frequency of the local ALP field is detectable, and thus cannot be used to explain the static isotropic birefringence signal.
We find that in the range $10^{-25}$~eV $\leq m_{\phi} < 10^{-23}$~eV, the upper limit of the coupling constant constrained by the CMB washout effect is a factor of 3 lower than the limit required to explain the cosmic birefringence effect.
Notably, due to the same scaling law of fraction parameter between Eq.~(\ref{eq:localbirefringence}) and Eq.~(\ref{eq:washout}), it is evident that there is no overlap between the two regardless of the value of $\kappa$, and any future observational advances in this mass range will not affect this conclusion.
Moreover, since ALPs with $m_{\phi} \gtrsim 10^{-23}$~eV cannot be used to explain static cosmic birefringence, we can conclude that any ALPs that can undergo nonlinear clustering in the late universe, and thus contributing mainly to the rotation angle of polarized photons, cannot be used to explain the static cosmic birefringence signal obtained from the analysis of Planck and WMAP data.

In Fig.~(\ref{fig:fullcons}), we have updated the $g_{\phi \gamma}-m_\phi$ relationship diagram compared to previous work \citep{2021PhRvD.103d3509F} explaining the best-fit static isotropic cosmic birefringence using ALPs.
ALPs act as dark energy when $m_\phi \lesssim 10^{-33}$ eV, and can act as a fraction of dark matter when $m_\phi \gtrsim 10^{-32}$ eV.
When $m_\phi \gtrsim 10^{-26}$ eV, ALPs gradually start to exhibit nonlinear clustering effects in the late universe. However, there are currently no precise simulations for calculations around mass $m_\phi \sim 10^{-26} \mathrm{eV}$, so we indicate this interval with a dashed line and leave it for future work.
Notably, due to the washout effect still being effective at $m_\phi \sim 10^{-26}$ eV, and the constraints on $g_{\phi \gamma}$ within this mass range being consistently stronger than the minimum required to explain cosmic birefringence (i.e., assuming that ALPs at $m_\phi \sim 10^{-26}$ eV still follow the CDM density distribution locally), this mass range still cannot be used to explain static isotropic cosmic birefringence.
We calculate the washout effect down to $10^{-26}$ eV, because by $10^{-27}$ eV, the visibility function would start to no longer satisfy $2\pi |\partial_t g(t)| \ll m_\phi$ in the LSS, and ALPs no longer have complex oscillations at the LSS. Therefore, the approximation in Eq.~(\ref{eq:ToyModel3Result2}) is no longer valid, and more precise calculation methods, such as accurate modifications to the Boltzmann equation, are required, which would be an interesting work in the future.
Furthermore, it is worth noting that Reference~\cite{2021PhRvD.103d3509F} did not consider the effect of birefringence tomography \cite{Sherwin:2021vgb,Nakatsuka:2022epj,Diego-Palazuelos:2024lym}, but this calculation is beyond the scope of this paper, and therefore Fig.~(\ref{fig:fullcons}) needs to be further corrected in future research.

\section{\label{Sec4}Conclusions}

In this study, we calculated the relationship between ALP mass and the coupling coefficient under the latest cosmological constraints, considering the CMB washout effect.
We studied the ADM mass range that can explain the static cosmic birefringence effect through local nonlinear evolution in the late universe, assuming that the birefringence contribution from the recombination period is canceled to a negligible level by the rapidly oscillating ALP field.
We found that the upper limit of the coupling coefficient, constrained by the CMB washout effect, is a factor of 3 lower than the limit needed to explain the cosmic birefringence effect.
This discrepancy rules out the proposed mechanism.
We briefly reviewed previous studies on ALP mass constraints and produced an updated ALP-coupling constant diagram.

We conclude that the mass range of ADM capable of nonlinear evolution in the late universe cannot explain the static cosmic birefringence phenomenon.
Therefore, if clustering ADM is considered a dark matter candidate and the static cosmic birefringence signal in CMB is confirmed with higher confidence in future surveys, further development of these models must introduce non-conventional potentials, multiple ALP fields or more complex theoretical frameworks.

\begin{acknowledgments}
We thank Yi-Fu Cai, Eiichiro Komatsu, Kai Murai and Tsutomu Yanagida for fruitful discussion and comments.
The Kavli IPMU is supported by the World Premier International Research Center Initiative (WPI), MEXT, Japan. 
DZ acknowledges the support from the China Scholarship Council.
EF thanks the support of the Serrapilheira Institute.
IO acknowledges the support from JSPS KAKENHI Grant No. JP19K14702.
TN acknowledges support from JSPS KAKENHI Grant No. JP20H05859 and No. JP22K03682.

\end{acknowledgments}

\bibliography{references}

\begin{thebibliography}{78}%
\makeatletter
\providecommand \@ifxundefined [1]{%
 \@ifx{#1\undefined}
}%
\providecommand \@ifnum [1]{%
 \ifnum #1\expandafter \@firstoftwo
 \else \expandafter \@secondoftwo
 \fi
}%
\providecommand \@ifx [1]{%
 \ifx #1\expandafter \@firstoftwo
 \else \expandafter \@secondoftwo
 \fi
}%
\providecommand \natexlab [1]{#1}%
\providecommand \enquote  [1]{``#1''}%
\providecommand \bibnamefont  [1]{#1}%
\providecommand \bibfnamefont [1]{#1}%
\providecommand \citenamefont [1]{#1}%
\providecommand \href@noop [0]{\@secondoftwo}%
\providecommand \href [0]{\begingroup \@sanitize@url \@href}%
\providecommand \@href[1]{\@@startlink{#1}\@@href}%
\providecommand \@@href[1]{\endgroup#1\@@endlink}%
\providecommand \@sanitize@url [0]{\catcode `\\12\catcode `\$12\catcode `\&12\catcode `\#12\catcode `\^12\catcode `\_12\catcode `\%12\relax}%
\providecommand \@@startlink[1]{}%
\providecommand \@@endlink[0]{}%
\providecommand \url  [0]{\begingroup\@sanitize@url \@url }%
\providecommand \@url [1]{\endgroup\@href {#1}{\urlprefix }}%
\providecommand \urlprefix  [0]{URL }%
\providecommand \Eprint [0]{\href }%
\providecommand \doibase [0]{https://doi.org/}%
\providecommand \selectlanguage [0]{\@gobble}%
\providecommand \bibinfo  [0]{\@secondoftwo}%
\providecommand \bibfield  [0]{\@secondoftwo}%
\providecommand \translation [1]{[#1]}%
\providecommand \BibitemOpen [0]{}%
\providecommand \bibitemStop [0]{}%
\providecommand \bibitemNoStop [0]{.\EOS\space}%
\providecommand \EOS [0]{\spacefactor3000\relax}%
\providecommand \BibitemShut  [1]{\csname bibitem#1\endcsname}%
\let\auto@bib@innerbib\@empty
%</preamble>
\bibitem [{\citenamefont {Minami}\ and\ \citenamefont {Komatsu}(2020)}]{2020PhRvL.125v1301M}%
  \BibitemOpen
  \bibfield  {author} {\bibinfo {author} {\bibfnamefont {Y.}~\bibnamefont {Minami}}\ and\ \bibinfo {author} {\bibfnamefont {E.}~\bibnamefont {Komatsu}},\ }\bibfield  {title} {\bibinfo {title} {{New Extraction of the Cosmic Birefringence from the Planck 2018 Polarization Data}},\ }\href {https://doi.org/10.1103/PhysRevLett.125.221301} {\bibfield  {journal} {\bibinfo  {journal} {Phys. Rev. Lett.}\ }\textbf {\bibinfo {volume} {125}},\ \bibinfo {pages} {221301} (\bibinfo {year} {2020})},\ \Eprint {https://arxiv.org/abs/2011.11254} {arXiv:2011.11254 [astro-ph.CO]} \BibitemShut {NoStop}%
\bibitem [{\citenamefont {Diego-Palazuelos}\ \emph {et~al.}(2022)\citenamefont {Diego-Palazuelos} \emph {et~al.}}]{2022PhRvL.128i1302D}%
  \BibitemOpen
  \bibfield  {author} {\bibinfo {author} {\bibfnamefont {P.}~\bibnamefont {Diego-Palazuelos}} \emph {et~al.},\ }\bibfield  {title} {\bibinfo {title} {{Cosmic Birefringence from the Planck Data Release 4}},\ }\href {https://doi.org/10.1103/PhysRevLett.128.091302} {\bibfield  {journal} {\bibinfo  {journal} {Phys. Rev. Lett.}\ }\textbf {\bibinfo {volume} {128}},\ \bibinfo {pages} {091302} (\bibinfo {year} {2022})},\ \Eprint {https://arxiv.org/abs/2201.07682} {arXiv:2201.07682 [astro-ph.CO]} \BibitemShut {NoStop}%
\bibitem [{\citenamefont {Eskilt}(2022)}]{2022A&A...662A..10E}%
  \BibitemOpen
  \bibfield  {author} {\bibinfo {author} {\bibfnamefont {J.~R.}\ \bibnamefont {Eskilt}},\ }\bibfield  {title} {\bibinfo {title} {{Frequency-dependent constraints on cosmic birefringence from the LFI and HFI Planck Data Release 4}},\ }\href {https://doi.org/10.1051/0004-6361/202243269} {\bibfield  {journal} {\bibinfo  {journal} {Astron. Astrophys.}\ }\textbf {\bibinfo {volume} {662}},\ \bibinfo {pages} {A10} (\bibinfo {year} {2022})},\ \Eprint {https://arxiv.org/abs/2201.13347} {arXiv:2201.13347 [astro-ph.CO]} \BibitemShut {NoStop}%
\bibitem [{\citenamefont {Eskilt}\ and\ \citenamefont {Komatsu}(2022)}]{2022PhRvD.106f3503E}%
  \BibitemOpen
  \bibfield  {author} {\bibinfo {author} {\bibfnamefont {J.~R.}\ \bibnamefont {Eskilt}}\ and\ \bibinfo {author} {\bibfnamefont {E.}~\bibnamefont {Komatsu}},\ }\bibfield  {title} {\bibinfo {title} {{Improved constraints on cosmic birefringence from the WMAP and Planck cosmic microwave background polarization data}},\ }\href {https://doi.org/10.1103/PhysRevD.106.063503} {\bibfield  {journal} {\bibinfo  {journal} {Phys. Rev. D}\ }\textbf {\bibinfo {volume} {106}},\ \bibinfo {pages} {063503} (\bibinfo {year} {2022})},\ \Eprint {https://arxiv.org/abs/2205.13962} {arXiv:2205.13962 [astro-ph.CO]} \BibitemShut {NoStop}%
\bibitem [{\citenamefont {Eskilt}\ \emph {et~al.}(2023{\natexlab{a}})\citenamefont {Eskilt} \emph {et~al.}}]{2023A&A...679A.144E}%
  \BibitemOpen
  \bibfield  {author} {\bibinfo {author} {\bibfnamefont {J.~R.}\ \bibnamefont {Eskilt}} \emph {et~al.} (\bibinfo {collaboration} {Cosmoglobe}),\ }\bibfield  {title} {\bibinfo {title} {{COSMOGLOBE DR1 results - II. Constraints on isotropic cosmic birefringence from reprocessed WMAP and Planck LFI data}},\ }\href {https://doi.org/10.1051/0004-6361/202346829} {\bibfield  {journal} {\bibinfo  {journal} {Astron. Astrophys.}\ }\textbf {\bibinfo {volume} {679}},\ \bibinfo {pages} {A144} (\bibinfo {year} {2023}{\natexlab{a}})},\ \Eprint {https://arxiv.org/abs/2305.02268} {arXiv:2305.02268 [astro-ph.CO]} \BibitemShut {NoStop}%
\bibitem [{\citenamefont {Komatsu}(2022)}]{2022NatRP...4..452K}%
  \BibitemOpen
  \bibfield  {author} {\bibinfo {author} {\bibfnamefont {E.}~\bibnamefont {Komatsu}},\ }\bibfield  {title} {\bibinfo {title} {{New physics from the polarized light of the cosmic microwave background}},\ }\href {https://doi.org/10.1038/s42254-022-00452-4} {\bibfield  {journal} {\bibinfo  {journal} {Nature Rev. Phys.}\ }\textbf {\bibinfo {volume} {4}},\ \bibinfo {pages} {452} (\bibinfo {year} {2022})},\ \Eprint {https://arxiv.org/abs/2202.13919} {arXiv:2202.13919 [astro-ph.CO]} \BibitemShut {NoStop}%
\bibitem [{\citenamefont {{BICEP2 and Keck Array Collaborations}}(2017)}]{BICEP2:2017lpa}%
  \BibitemOpen
  \bibfield  {author} {\bibinfo {author} {\bibnamefont {{BICEP2 and Keck Array Collaborations}}},\ }\bibfield  {title} {\bibinfo {title} {{BICEP2 / Keck Array IX: New bounds on anisotropies of CMB polarization rotation and implications for axionlike particles and primordial magnetic fields}},\ }\href {https://doi.org/10.1103/PhysRevD.96.102003} {\bibfield  {journal} {\bibinfo  {journal} {\prd}\ }\textbf {\bibinfo {volume} {96}},\ \bibinfo {pages} {102003} (\bibinfo {year} {2017})},\ \Eprint {https://arxiv.org/abs/1705.02523} {1705.02523} \BibitemShut {NoStop}%
\bibitem [{\citenamefont {{BICEP2 and Keck Array Collaborations}}(2023)}]{BK-LoS:2023}%
  \BibitemOpen
  \bibfield  {author} {\bibinfo {author} {\bibnamefont {{BICEP2 and Keck Array Collaborations}}},\ }\bibfield  {title} {\bibinfo {title} {{BICEP/Keck. XVII. Line-of-sight Distortion Analysis: Estimates of Gravitational Lensing, Anisotropic Cosmic Birefringence, Patchy Reionization, and Systematic Errors}},\ }\href {https://doi.org/10.3847/1538-4357/acc85c} {\bibfield  {journal} {\bibinfo  {journal} {\apj}\ }\textbf {\bibinfo {volume} {949}},\ \bibinfo {pages} {43} (\bibinfo {year} {2023})},\ \Eprint {https://arxiv.org/abs/2210.08038} {2210.08038} \BibitemShut {NoStop}%
\bibitem [{\citenamefont {Namikawa}\ \emph {et~al.}(2020)\citenamefont {Namikawa} \emph {et~al.}}]{2020PhRvD.101h3527N}%
  \BibitemOpen
  \bibfield  {author} {\bibinfo {author} {\bibfnamefont {T.}~\bibnamefont {Namikawa}} \emph {et~al.},\ }\bibfield  {title} {\bibinfo {title} {{Atacama Cosmology Telescope: Constraints on cosmic birefringence}},\ }\href {https://doi.org/10.1103/PhysRevD.101.083527} {\bibfield  {journal} {\bibinfo  {journal} {Phys. Rev. D}\ }\textbf {\bibinfo {volume} {101}},\ \bibinfo {pages} {083527} (\bibinfo {year} {2020})},\ \Eprint {https://arxiv.org/abs/2001.10465} {arXiv:2001.10465 [astro-ph.CO]} \BibitemShut {NoStop}%
\bibitem [{\citenamefont {Bianchini}\ \emph {et~al.}(2020)\citenamefont {Bianchini} \emph {et~al.}}]{2020PhRvD.102h3504B}%
  \BibitemOpen
  \bibfield  {author} {\bibinfo {author} {\bibfnamefont {F.}~\bibnamefont {Bianchini}} \emph {et~al.} (\bibinfo {collaboration} {SPT}),\ }\bibfield  {title} {\bibinfo {title} {{Searching for Anisotropic Cosmic Birefringence with Polarization Data from SPTpol}},\ }\href {https://doi.org/10.1103/PhysRevD.102.083504} {\bibfield  {journal} {\bibinfo  {journal} {Phys. Rev. D}\ }\textbf {\bibinfo {volume} {102}},\ \bibinfo {pages} {083504} (\bibinfo {year} {2020})},\ \Eprint {https://arxiv.org/abs/2006.08061} {arXiv:2006.08061 [astro-ph.CO]} \BibitemShut {NoStop}%
\bibitem [{\citenamefont {Bortolami}\ \emph {et~al.}(2022)\citenamefont {Bortolami}, \citenamefont {Billi}, \citenamefont {Gruppuso}, \citenamefont {Natoli},\ and\ \citenamefont {Pagano}}]{Bortolami:2022whx}%
  \BibitemOpen
  \bibfield  {author} {\bibinfo {author} {\bibfnamefont {M.}~\bibnamefont {Bortolami}}, \bibinfo {author} {\bibfnamefont {M.}~\bibnamefont {Billi}}, \bibinfo {author} {\bibfnamefont {A.}~\bibnamefont {Gruppuso}}, \bibinfo {author} {\bibfnamefont {P.}~\bibnamefont {Natoli}},\ and\ \bibinfo {author} {\bibfnamefont {L.}~\bibnamefont {Pagano}},\ }\bibfield  {title} {\bibinfo {title} {{Planck constraints on cross-correlations between anisotropic cosmic birefringence and CMB polarization}},\ }\href {https://doi.org/10.1088/1475-7516/2022/09/075} {\bibfield  {journal} {\bibinfo  {journal} {\jcap}\ }\textbf {\bibinfo {volume} {09}},\ \bibinfo {pages} {075} (\bibinfo {year} {2022})},\ \Eprint {https://arxiv.org/abs/2206.01635} {2206.01635} \BibitemShut {NoStop}%
\bibitem [{\citenamefont {Zagatti}\ \emph {et~al.}(2024)\citenamefont {Zagatti}, \citenamefont {Bortolami}, \citenamefont {Gruppuso}, \citenamefont {Natoli}, \citenamefont {Pagano},\ and\ \citenamefont {Fabbian}}]{Zagatti:2024}%
  \BibitemOpen
  \bibfield  {author} {\bibinfo {author} {\bibfnamefont {G.}~\bibnamefont {Zagatti}}, \bibinfo {author} {\bibfnamefont {M.}~\bibnamefont {Bortolami}}, \bibinfo {author} {\bibfnamefont {A.}~\bibnamefont {Gruppuso}}, \bibinfo {author} {\bibfnamefont {P.}~\bibnamefont {Natoli}}, \bibinfo {author} {\bibfnamefont {L.}~\bibnamefont {Pagano}},\ and\ \bibinfo {author} {\bibfnamefont {G.}~\bibnamefont {Fabbian}},\ }\bibfield  {title} {\bibinfo {title} {{Planck constraints on Cosmic Birefringence and its cross-correlation with the CMB}},\ }\href@noop {} {\  (\bibinfo {year} {2024})},\ \Eprint {https://arxiv.org/abs/2401.11973} {2401.11973} \BibitemShut {NoStop}%
\bibitem [{\citenamefont {Namikawa}(2024)}]{Namikawa:2024:BB}%
  \BibitemOpen
  \bibfield  {author} {\bibinfo {author} {\bibfnamefont {T.}~\bibnamefont {Namikawa}},\ }\bibfield  {title} {\bibinfo {title} {{Exact CMB B-mode power spectrum from anisotropic cosmic birefringence}},\ }\href@noop {} {\bibfield  {journal} {\bibinfo  {journal} {\prd}\ }\textbf {\bibinfo {volume} {109}},\ \bibinfo {pages} {123521} (\bibinfo {year} {2024})},\ \Eprint {https://arxiv.org/abs/2404.13771} {2404.13771} \BibitemShut {NoStop}%
\bibitem [{\citenamefont {Adachi}\ \emph {et~al.}(2024)\citenamefont {Adachi} \emph {et~al.}}]{2024arXiv240302096A}%
  \BibitemOpen
  \bibfield  {author} {\bibinfo {author} {\bibfnamefont {S.}~\bibnamefont {Adachi}} \emph {et~al.},\ }\bibfield  {title} {\bibinfo {title} {Exploration of the polarization angle variability of the crab nebula with polarbear and its application to the search for axion-like particles},\ }\href@noop {} {\bibfield  {journal} {\bibinfo  {journal} {arXiv preprint arXiv:2403.02096}\ } (\bibinfo {year} {2024})}\BibitemShut {NoStop}%
\bibitem [{\citenamefont {Fujita}\ \emph {et~al.}(2021{\natexlab{a}})\citenamefont {Fujita}, \citenamefont {Minami}, \citenamefont {Murai},\ and\ \citenamefont {Nakatsuka}}]{2020arXiv200802473F}%
  \BibitemOpen
  \bibfield  {author} {\bibinfo {author} {\bibfnamefont {T.}~\bibnamefont {Fujita}}, \bibinfo {author} {\bibfnamefont {Y.}~\bibnamefont {Minami}}, \bibinfo {author} {\bibfnamefont {K.}~\bibnamefont {Murai}},\ and\ \bibinfo {author} {\bibfnamefont {H.}~\bibnamefont {Nakatsuka}},\ }\bibfield  {title} {\bibinfo {title} {{Probing axionlike particles via cosmic microwave background polarization}},\ }\href {https://doi.org/10.1103/PhysRevD.103.063508} {\bibfield  {journal} {\bibinfo  {journal} {Phys. Rev. D}\ }\textbf {\bibinfo {volume} {103}},\ \bibinfo {pages} {063508} (\bibinfo {year} {2021}{\natexlab{a}})},\ \Eprint {https://arxiv.org/abs/2008.02473} {arXiv:2008.02473 [astro-ph.CO]} \BibitemShut {NoStop}%
\bibitem [{\citenamefont {Harari}\ and\ \citenamefont {Sikivie}(1992)}]{Harari:1992ea}%
  \BibitemOpen
  \bibfield  {author} {\bibinfo {author} {\bibfnamefont {D.}~\bibnamefont {Harari}}\ and\ \bibinfo {author} {\bibfnamefont {P.}~\bibnamefont {Sikivie}},\ }\bibfield  {title} {\bibinfo {title} {{Effects of a Nambu-Goldstone boson on the polarization of radio galaxies and the cosmic microwave background}},\ }\href {https://doi.org/10.1016/0370-2693(92)91363-E} {\bibfield  {journal} {\bibinfo  {journal} {Phys. Lett. B}\ }\textbf {\bibinfo {volume} {289}},\ \bibinfo {pages} {67} (\bibinfo {year} {1992})}\BibitemShut {NoStop}%
\bibitem [{\citenamefont {Carroll}(1998)}]{1998PhRvL..81.3067C}%
  \BibitemOpen
  \bibfield  {author} {\bibinfo {author} {\bibfnamefont {S.~M.}\ \bibnamefont {Carroll}},\ }\bibfield  {title} {\bibinfo {title} {{Quintessence and the rest of the world}},\ }\href {https://doi.org/10.1103/PhysRevLett.81.3067} {\bibfield  {journal} {\bibinfo  {journal} {Phys. Rev. Lett.}\ }\textbf {\bibinfo {volume} {81}},\ \bibinfo {pages} {3067} (\bibinfo {year} {1998})},\ \Eprint {https://arxiv.org/abs/astro-ph/9806099} {arXiv:astro-ph/9806099} \BibitemShut {NoStop}%
\bibitem [{\citenamefont {Panda}\ \emph {et~al.}(2011)\citenamefont {Panda}, \citenamefont {Sumitomo},\ and\ \citenamefont {Trivedi}}]{2011PhRvD..83h3506P}%
  \BibitemOpen
  \bibfield  {author} {\bibinfo {author} {\bibfnamefont {S.}~\bibnamefont {Panda}}, \bibinfo {author} {\bibfnamefont {Y.}~\bibnamefont {Sumitomo}},\ and\ \bibinfo {author} {\bibfnamefont {S.~P.}\ \bibnamefont {Trivedi}},\ }\bibfield  {title} {\bibinfo {title} {{Axions as Quintessence in String Theory}},\ }\href {https://doi.org/10.1103/PhysRevD.83.083506} {\bibfield  {journal} {\bibinfo  {journal} {Phys. Rev. D}\ }\textbf {\bibinfo {volume} {83}},\ \bibinfo {pages} {083506} (\bibinfo {year} {2011})},\ \Eprint {https://arxiv.org/abs/1011.5877} {arXiv:1011.5877 [hep-th]} \BibitemShut {NoStop}%
\bibitem [{\citenamefont {Fujita}\ \emph {et~al.}(2021{\natexlab{b}})\citenamefont {Fujita}, \citenamefont {Murai}, \citenamefont {Nakatsuka},\ and\ \citenamefont {Tsujikawa}}]{2021PhRvD.103d3509F}%
  \BibitemOpen
  \bibfield  {author} {\bibinfo {author} {\bibfnamefont {T.}~\bibnamefont {Fujita}}, \bibinfo {author} {\bibfnamefont {K.}~\bibnamefont {Murai}}, \bibinfo {author} {\bibfnamefont {H.}~\bibnamefont {Nakatsuka}},\ and\ \bibinfo {author} {\bibfnamefont {S.}~\bibnamefont {Tsujikawa}},\ }\bibfield  {title} {\bibinfo {title} {{Detection of isotropic cosmic birefringence and its implications for axionlike particles including dark energy}},\ }\href {https://doi.org/10.1103/PhysRevD.103.043509} {\bibfield  {journal} {\bibinfo  {journal} {Phys. Rev. D}\ }\textbf {\bibinfo {volume} {103}},\ \bibinfo {pages} {043509} (\bibinfo {year} {2021}{\natexlab{b}})},\ \Eprint {https://arxiv.org/abs/2011.11894} {arXiv:2011.11894 [astro-ph.CO]} \BibitemShut {NoStop}%
\bibitem [{\citenamefont {Choi}\ \emph {et~al.}(2021)\citenamefont {Choi}, \citenamefont {Lin}, \citenamefont {Visinelli},\ and\ \citenamefont {Yanagida}}]{2021PhRvD.104j1302C}%
  \BibitemOpen
  \bibfield  {author} {\bibinfo {author} {\bibfnamefont {G.}~\bibnamefont {Choi}}, \bibinfo {author} {\bibfnamefont {W.}~\bibnamefont {Lin}}, \bibinfo {author} {\bibfnamefont {L.}~\bibnamefont {Visinelli}},\ and\ \bibinfo {author} {\bibfnamefont {T.~T.}\ \bibnamefont {Yanagida}},\ }\bibfield  {title} {\bibinfo {title} {{Cosmic birefringence and electroweak axion dark energy}},\ }\href {https://doi.org/10.1103/PhysRevD.104.L101302} {\bibfield  {journal} {\bibinfo  {journal} {Phys. Rev. D}\ }\textbf {\bibinfo {volume} {104}},\ \bibinfo {pages} {L101302} (\bibinfo {year} {2021})},\ \Eprint {https://arxiv.org/abs/2106.12602} {arXiv:2106.12602 [hep-ph]} \BibitemShut {NoStop}%
\bibitem [{\citenamefont {Obata}(2022)}]{2022JCAP...09..062O}%
  \BibitemOpen
  \bibfield  {author} {\bibinfo {author} {\bibfnamefont {I.}~\bibnamefont {Obata}},\ }\bibfield  {title} {\bibinfo {title} {{Implications of the cosmic birefringence measurement for the axion dark matter search}},\ }\href {https://doi.org/10.1088/1475-7516/2022/09/062} {\bibfield  {journal} {\bibinfo  {journal} {JCAP}\ }\textbf {\bibinfo {volume} {09}},\ \bibinfo {pages} {062}},\ \Eprint {https://arxiv.org/abs/2108.02150} {arXiv:2108.02150 [astro-ph.CO]} \BibitemShut {NoStop}%
\bibitem [{\citenamefont {Gasparotto}\ and\ \citenamefont {Obata}(2022)}]{2022JCAP...08..025G}%
  \BibitemOpen
  \bibfield  {author} {\bibinfo {author} {\bibfnamefont {S.}~\bibnamefont {Gasparotto}}\ and\ \bibinfo {author} {\bibfnamefont {I.}~\bibnamefont {Obata}},\ }\bibfield  {title} {\bibinfo {title} {{Cosmic birefringence from monodromic axion dark energy}},\ }\href {https://doi.org/10.1088/1475-7516/2022/08/025} {\bibfield  {journal} {\bibinfo  {journal} {JCAP}\ }\textbf {\bibinfo {volume} {08}}\bibfield  {number} {\bibinfo  {number} { (08)},\ \bibinfo {pages} {025}},\ }\Eprint {https://arxiv.org/abs/2203.09409} {arXiv:2203.09409 [astro-ph.CO]} \BibitemShut {NoStop}%
\bibitem [{\citenamefont {Murai}\ \emph {et~al.}(2023)\citenamefont {Murai}, \citenamefont {Naokawa}, \citenamefont {Namikawa},\ and\ \citenamefont {Komatsu}}]{2023PhRvD.107d1302M}%
  \BibitemOpen
  \bibfield  {author} {\bibinfo {author} {\bibfnamefont {K.}~\bibnamefont {Murai}}, \bibinfo {author} {\bibfnamefont {F.}~\bibnamefont {Naokawa}}, \bibinfo {author} {\bibfnamefont {T.}~\bibnamefont {Namikawa}},\ and\ \bibinfo {author} {\bibfnamefont {E.}~\bibnamefont {Komatsu}},\ }\bibfield  {title} {\bibinfo {title} {{Isotropic cosmic birefringence from early dark energy}},\ }\href {https://doi.org/10.1103/PhysRevD.107.L041302} {\bibfield  {journal} {\bibinfo  {journal} {Phys. Rev. D}\ }\textbf {\bibinfo {volume} {107}},\ \bibinfo {pages} {L041302} (\bibinfo {year} {2023})},\ \Eprint {https://arxiv.org/abs/2209.07804} {arXiv:2209.07804 [astro-ph.CO]} \BibitemShut {NoStop}%
\bibitem [{\citenamefont {Eskilt}\ \emph {et~al.}(2023{\natexlab{b}})\citenamefont {Eskilt}, \citenamefont {Herold}, \citenamefont {Komatsu}, \citenamefont {Murai}, \citenamefont {Namikawa},\ and\ \citenamefont {Naokawa}}]{2023PhRvL.131l1001E}%
  \BibitemOpen
  \bibfield  {author} {\bibinfo {author} {\bibfnamefont {J.~R.}\ \bibnamefont {Eskilt}}, \bibinfo {author} {\bibfnamefont {L.}~\bibnamefont {Herold}}, \bibinfo {author} {\bibfnamefont {E.}~\bibnamefont {Komatsu}}, \bibinfo {author} {\bibfnamefont {K.}~\bibnamefont {Murai}}, \bibinfo {author} {\bibfnamefont {T.}~\bibnamefont {Namikawa}},\ and\ \bibinfo {author} {\bibfnamefont {F.}~\bibnamefont {Naokawa}},\ }\bibfield  {title} {\bibinfo {title} {{Constraints on Early Dark Energy from Isotropic Cosmic Birefringence}},\ }\href {https://doi.org/10.1103/PhysRevLett.131.121001} {\bibfield  {journal} {\bibinfo  {journal} {Phys. Rev. Lett.}\ }\textbf {\bibinfo {volume} {131}},\ \bibinfo {pages} {121001} (\bibinfo {year} {2023}{\natexlab{b}})},\ \Eprint {https://arxiv.org/abs/2303.15369} {arXiv:2303.15369 [astro-ph.CO]} \BibitemShut {NoStop}%
\bibitem [{\citenamefont {Takahashi}\ and\ \citenamefont {Yin}(2021)}]{Takahashi:2020tqv}%
  \BibitemOpen
  \bibfield  {author} {\bibinfo {author} {\bibfnamefont {F.}~\bibnamefont {Takahashi}}\ and\ \bibinfo {author} {\bibfnamefont {W.}~\bibnamefont {Yin}},\ }\bibfield  {title} {\bibinfo {title} {{Kilobyte Cosmic Birefringence from ALP Domain Walls}},\ }\href {https://doi.org/10.1088/1475-7516/2021/04/007} {\bibfield  {journal} {\bibinfo  {journal} {JCAP}\ }\textbf {\bibinfo {volume} {04}},\ \bibinfo {pages} {007}},\ \Eprint {https://arxiv.org/abs/2012.11576} {arXiv:2012.11576 [hep-ph]} \BibitemShut {NoStop}%
\bibitem [{\citenamefont {Kitajima}\ \emph {et~al.}(2022)\citenamefont {Kitajima}, \citenamefont {Kozai}, \citenamefont {Takahashi},\ and\ \citenamefont {Yin}}]{Kitajima:2022jzz}%
  \BibitemOpen
  \bibfield  {author} {\bibinfo {author} {\bibfnamefont {N.}~\bibnamefont {Kitajima}}, \bibinfo {author} {\bibfnamefont {F.}~\bibnamefont {Kozai}}, \bibinfo {author} {\bibfnamefont {F.}~\bibnamefont {Takahashi}},\ and\ \bibinfo {author} {\bibfnamefont {W.}~\bibnamefont {Yin}},\ }\bibfield  {title} {\bibinfo {title} {{Power spectrum of domain-wall network, and its implications for isotropic and anisotropic cosmic birefringence}},\ }\href {https://doi.org/10.1088/1475-7516/2022/10/043} {\bibfield  {journal} {\bibinfo  {journal} {JCAP}\ }\textbf {\bibinfo {volume} {10}},\ \bibinfo {pages} {043}},\ \Eprint {https://arxiv.org/abs/2205.05083} {arXiv:2205.05083 [astro-ph.CO]} \BibitemShut {NoStop}%
\bibitem [{\citenamefont {Gonzalez}\ \emph {et~al.}(2023)\citenamefont {Gonzalez}, \citenamefont {Kitajima}, \citenamefont {Takahashi},\ and\ \citenamefont {Yin}}]{Gonzalez:2022mcx}%
  \BibitemOpen
  \bibfield  {author} {\bibinfo {author} {\bibfnamefont {D.}~\bibnamefont {Gonzalez}}, \bibinfo {author} {\bibfnamefont {N.}~\bibnamefont {Kitajima}}, \bibinfo {author} {\bibfnamefont {F.}~\bibnamefont {Takahashi}},\ and\ \bibinfo {author} {\bibfnamefont {W.}~\bibnamefont {Yin}},\ }\bibfield  {title} {\bibinfo {title} {{Stability of domain wall network with initial inflationary fluctuations and its implications for cosmic birefringence}},\ }\href {https://doi.org/10.1016/j.physletb.2023.137990} {\bibfield  {journal} {\bibinfo  {journal} {Phys. Lett. B}\ }\textbf {\bibinfo {volume} {843}},\ \bibinfo {pages} {137990} (\bibinfo {year} {2023})},\ \Eprint {https://arxiv.org/abs/2211.06849} {arXiv:2211.06849 [hep-ph]} \BibitemShut {NoStop}%
\bibitem [{\citenamefont {Kitajima}\ \emph {et~al.}(2023)\citenamefont {Kitajima}, \citenamefont {Lee}, \citenamefont {Takahashi},\ and\ \citenamefont {Yin}}]{Kitajima:2023kzu}%
  \BibitemOpen
  \bibfield  {author} {\bibinfo {author} {\bibfnamefont {N.}~\bibnamefont {Kitajima}}, \bibinfo {author} {\bibfnamefont {J.}~\bibnamefont {Lee}}, \bibinfo {author} {\bibfnamefont {F.}~\bibnamefont {Takahashi}},\ and\ \bibinfo {author} {\bibfnamefont {W.}~\bibnamefont {Yin}},\ }\bibfield  {title} {\bibinfo {title} {Stability of domain walls with inflationary fluctuations under potential bias, and gravitational wave signatures},\ }\href@noop {} {\bibfield  {journal} {\bibinfo  {journal} {arXiv preprint arXiv:2311.14590}\ } (\bibinfo {year} {2023})}\BibitemShut {NoStop}%
\bibitem [{\citenamefont {Ferreira}\ \emph {et~al.}(2024)\citenamefont {Ferreira}, \citenamefont {Gasparotto}, \citenamefont {Hiramatsu}, \citenamefont {Obata},\ and\ \citenamefont {Pujolas}}]{Ferreira:2023jbu}%
  \BibitemOpen
  \bibfield  {author} {\bibinfo {author} {\bibfnamefont {R.~Z.}\ \bibnamefont {Ferreira}}, \bibinfo {author} {\bibfnamefont {S.}~\bibnamefont {Gasparotto}}, \bibinfo {author} {\bibfnamefont {T.}~\bibnamefont {Hiramatsu}}, \bibinfo {author} {\bibfnamefont {I.}~\bibnamefont {Obata}},\ and\ \bibinfo {author} {\bibfnamefont {O.}~\bibnamefont {Pujolas}},\ }\bibfield  {title} {\bibinfo {title} {{Axionic defects in the CMB: birefringence and gravitational waves}},\ }\href {https://doi.org/10.1088/1475-7516/2024/05/066} {\bibfield  {journal} {\bibinfo  {journal} {JCAP}\ }\textbf {\bibinfo {volume} {05}},\ \bibinfo {pages} {066}},\ \Eprint {https://arxiv.org/abs/2312.14104} {arXiv:2312.14104 [hep-ph]} \BibitemShut {NoStop}%
\bibitem [{\citenamefont {Finelli}\ and\ \citenamefont {Galaverni}(2009)}]{2009PhRvD..79f3002F}%
  \BibitemOpen
  \bibfield  {author} {\bibinfo {author} {\bibfnamefont {F.}~\bibnamefont {Finelli}}\ and\ \bibinfo {author} {\bibfnamefont {M.}~\bibnamefont {Galaverni}},\ }\bibfield  {title} {\bibinfo {title} {{Rotation of Linear Polarization Plane and Circular Polarization from Cosmological Pseudo-Scalar Fields}},\ }\href {https://doi.org/10.1103/PhysRevD.79.063002} {\bibfield  {journal} {\bibinfo  {journal} {Phys. Rev. D}\ }\textbf {\bibinfo {volume} {79}},\ \bibinfo {pages} {063002} (\bibinfo {year} {2009})},\ \Eprint {https://arxiv.org/abs/0802.4210} {arXiv:0802.4210 [astro-ph]} \BibitemShut {NoStop}%
\bibitem [{\citenamefont {Liu}\ and\ \citenamefont {Ng}(2017)}]{2017PDU....16...22L}%
  \BibitemOpen
  \bibfield  {author} {\bibinfo {author} {\bibfnamefont {G.-C.}\ \bibnamefont {Liu}}\ and\ \bibinfo {author} {\bibfnamefont {K.-W.}\ \bibnamefont {Ng}},\ }\bibfield  {title} {\bibinfo {title} {{Axion Dark Matter Induced Cosmic Microwave Background $B$-modes}},\ }\href {https://doi.org/10.1016/j.dark.2017.02.004} {\bibfield  {journal} {\bibinfo  {journal} {Phys. Dark Univ.}\ }\textbf {\bibinfo {volume} {16}},\ \bibinfo {pages} {22} (\bibinfo {year} {2017})},\ \Eprint {https://arxiv.org/abs/1612.02104} {arXiv:1612.02104 [astro-ph.CO]} \BibitemShut {NoStop}%
\bibitem [{\citenamefont {Fedderke}\ \emph {et~al.}(2019)\citenamefont {Fedderke}, \citenamefont {Graham},\ and\ \citenamefont {Rajendran}}]{2019PhRvD.100a5040F}%
  \BibitemOpen
  \bibfield  {author} {\bibinfo {author} {\bibfnamefont {M.~A.}\ \bibnamefont {Fedderke}}, \bibinfo {author} {\bibfnamefont {P.~W.}\ \bibnamefont {Graham}},\ and\ \bibinfo {author} {\bibfnamefont {S.}~\bibnamefont {Rajendran}},\ }\bibfield  {title} {\bibinfo {title} {{Axion Dark Matter Detection with CMB Polarization}},\ }\href {https://doi.org/10.1103/PhysRevD.100.015040} {\bibfield  {journal} {\bibinfo  {journal} {Phys. Rev. D}\ }\textbf {\bibinfo {volume} {100}},\ \bibinfo {pages} {015040} (\bibinfo {year} {2019})},\ \Eprint {https://arxiv.org/abs/1903.02666} {arXiv:1903.02666 [astro-ph.CO]} \BibitemShut {NoStop}%
\bibitem [{\citenamefont {Murai}(2024)}]{Murai:2024yul}%
  \BibitemOpen
  \bibfield  {author} {\bibinfo {author} {\bibfnamefont {K.}~\bibnamefont {Murai}},\ }\bibfield  {title} {\bibinfo {title} {Isotropic cosmic birefringence from an oscillating axion-like field},\ }\href@noop {} {\bibfield  {journal} {\bibinfo  {journal} {arXiv preprint arXiv:2407.14162}\ } (\bibinfo {year} {2024})}\BibitemShut {NoStop}%
\bibitem [{\citenamefont {Sherwin}\ and\ \citenamefont {Namikawa}(2023)}]{Sherwin:2021vgb}%
  \BibitemOpen
  \bibfield  {author} {\bibinfo {author} {\bibfnamefont {B.~D.}\ \bibnamefont {Sherwin}}\ and\ \bibinfo {author} {\bibfnamefont {T.}~\bibnamefont {Namikawa}},\ }\bibfield  {title} {\bibinfo {title} {{Cosmic birefringence tomography and calibration independence with reionization signals in the CMB}},\ }\href {https://doi.org/10.1093/mnras/stac3146} {\bibfield  {journal} {\bibinfo  {journal} {Mon. Not. Roy. Astron. Soc.}\ }\textbf {\bibinfo {volume} {520}},\ \bibinfo {pages} {3298} (\bibinfo {year} {2023})},\ \Eprint {https://arxiv.org/abs/2108.09287} {arXiv:2108.09287 [astro-ph.CO]} \BibitemShut {NoStop}%
\bibitem [{\citenamefont {Nakatsuka}\ \emph {et~al.}(2022)\citenamefont {Nakatsuka}, \citenamefont {Namikawa},\ and\ \citenamefont {Komatsu}}]{Nakatsuka:2022epj}%
  \BibitemOpen
  \bibfield  {author} {\bibinfo {author} {\bibfnamefont {H.}~\bibnamefont {Nakatsuka}}, \bibinfo {author} {\bibfnamefont {T.}~\bibnamefont {Namikawa}},\ and\ \bibinfo {author} {\bibfnamefont {E.}~\bibnamefont {Komatsu}},\ }\bibfield  {title} {\bibinfo {title} {{Is cosmic birefringence due to dark energy or dark matter? A tomographic approach}},\ }\href {https://doi.org/10.1103/PhysRevD.105.123509} {\bibfield  {journal} {\bibinfo  {journal} {Phys. Rev. D}\ }\textbf {\bibinfo {volume} {105}},\ \bibinfo {pages} {123509} (\bibinfo {year} {2022})},\ \Eprint {https://arxiv.org/abs/2203.08560} {arXiv:2203.08560 [astro-ph.CO]} \BibitemShut {NoStop}%
\bibitem [{\citenamefont {Lee}\ \emph {et~al.}(2022)\citenamefont {Lee}, \citenamefont {Hotinli},\ and\ \citenamefont {Kamionkowski}}]{Lee:2022udm}%
  \BibitemOpen
  \bibfield  {author} {\bibinfo {author} {\bibfnamefont {N.}~\bibnamefont {Lee}}, \bibinfo {author} {\bibfnamefont {S.~C.}\ \bibnamefont {Hotinli}},\ and\ \bibinfo {author} {\bibfnamefont {M.}~\bibnamefont {Kamionkowski}},\ }\bibfield  {title} {\bibinfo {title} {{Probing cosmic birefringence with polarized Sunyaev-Zel\textquoteright{}dovich tomography}},\ }\href {https://doi.org/10.1103/PhysRevD.106.083518} {\bibfield  {journal} {\bibinfo  {journal} {Phys. Rev. D}\ }\textbf {\bibinfo {volume} {106}},\ \bibinfo {pages} {083518} (\bibinfo {year} {2022})},\ \Eprint {https://arxiv.org/abs/2207.05687} {arXiv:2207.05687 [astro-ph.CO]} \BibitemShut {NoStop}%
\bibitem [{\citenamefont {Galaverni}\ \emph {et~al.}(2023)\citenamefont {Galaverni}, \citenamefont {Finelli},\ and\ \citenamefont {Paoletti}}]{2023PhRvD.107h3529G}%
  \BibitemOpen
  \bibfield  {author} {\bibinfo {author} {\bibfnamefont {M.}~\bibnamefont {Galaverni}}, \bibinfo {author} {\bibfnamefont {F.}~\bibnamefont {Finelli}},\ and\ \bibinfo {author} {\bibfnamefont {D.}~\bibnamefont {Paoletti}},\ }\bibfield  {title} {\bibinfo {title} {{Redshift evolution of cosmic birefringence in CMB anisotropies}},\ }\href {https://doi.org/10.1103/PhysRevD.107.083529} {\bibfield  {journal} {\bibinfo  {journal} {Phys. Rev. D}\ }\textbf {\bibinfo {volume} {107}},\ \bibinfo {pages} {083529} (\bibinfo {year} {2023})},\ \Eprint {https://arxiv.org/abs/2301.07971} {arXiv:2301.07971 [astro-ph.CO]} \BibitemShut {NoStop}%
\bibitem [{\citenamefont {Naokawa}\ and\ \citenamefont {Namikawa}(2023)}]{Naokawa:2023upt}%
  \BibitemOpen
  \bibfield  {author} {\bibinfo {author} {\bibfnamefont {F.}~\bibnamefont {Naokawa}}\ and\ \bibinfo {author} {\bibfnamefont {T.}~\bibnamefont {Namikawa}},\ }\bibfield  {title} {\bibinfo {title} {{Gravitational lensing effect on cosmic birefringence}},\ }\href {https://doi.org/10.1103/PhysRevD.108.063525} {\bibfield  {journal} {\bibinfo  {journal} {Phys. Rev. D}\ }\textbf {\bibinfo {volume} {108}},\ \bibinfo {pages} {063525} (\bibinfo {year} {2023})},\ \Eprint {https://arxiv.org/abs/2305.13976} {arXiv:2305.13976 [astro-ph.CO]} \BibitemShut {NoStop}%
\bibitem [{\citenamefont {Namikawa}\ and\ \citenamefont {Obata}(2023)}]{Namikawa:2023zux}%
  \BibitemOpen
  \bibfield  {author} {\bibinfo {author} {\bibfnamefont {T.}~\bibnamefont {Namikawa}}\ and\ \bibinfo {author} {\bibfnamefont {I.}~\bibnamefont {Obata}},\ }\bibfield  {title} {\bibinfo {title} {{Cosmic birefringence tomography with polarized Sunyaev-Zel\textquoteright{}dovich effect}},\ }\href {https://doi.org/10.1103/PhysRevD.108.083510} {\bibfield  {journal} {\bibinfo  {journal} {Phys. Rev. D}\ }\textbf {\bibinfo {volume} {108}},\ \bibinfo {pages} {083510} (\bibinfo {year} {2023})},\ \Eprint {https://arxiv.org/abs/2306.08875} {arXiv:2306.08875 [astro-ph.CO]} \BibitemShut {NoStop}%
\bibitem [{\citenamefont {Naokawa}\ \emph {et~al.}(2024)\citenamefont {Naokawa}, \citenamefont {Namikawa}, \citenamefont {Murai}, \citenamefont {Obata},\ and\ \citenamefont {Kamada}}]{Naokawa:2024xhn}%
  \BibitemOpen
  \bibfield  {author} {\bibinfo {author} {\bibfnamefont {F.}~\bibnamefont {Naokawa}}, \bibinfo {author} {\bibfnamefont {T.}~\bibnamefont {Namikawa}}, \bibinfo {author} {\bibfnamefont {K.}~\bibnamefont {Murai}}, \bibinfo {author} {\bibfnamefont {I.}~\bibnamefont {Obata}},\ and\ \bibinfo {author} {\bibfnamefont {K.}~\bibnamefont {Kamada}},\ }\bibfield  {title} {\bibinfo {title} {$n\pi$ phase ambiguity of cosmic birefringence},\ }\href@noop {} {\bibfield  {journal} {\bibinfo  {journal} {arXiv preprint arXiv:2405.15538}\ } (\bibinfo {year} {2024})}\BibitemShut {NoStop}%
\bibitem [{\citenamefont {Nakai}\ \emph {et~al.}(2024)\citenamefont {Nakai}, \citenamefont {Namba}, \citenamefont {Obata}, \citenamefont {Qiu},\ and\ \citenamefont {Saito}}]{2024JHEP...01..057N}%
  \BibitemOpen
  \bibfield  {author} {\bibinfo {author} {\bibfnamefont {Y.}~\bibnamefont {Nakai}}, \bibinfo {author} {\bibfnamefont {R.}~\bibnamefont {Namba}}, \bibinfo {author} {\bibfnamefont {I.}~\bibnamefont {Obata}}, \bibinfo {author} {\bibfnamefont {Y.-C.}\ \bibnamefont {Qiu}},\ and\ \bibinfo {author} {\bibfnamefont {R.}~\bibnamefont {Saito}},\ }\bibfield  {title} {\bibinfo {title} {{Can we explain cosmic birefringence without a new light field beyond Standard Model?}},\ }\href {https://doi.org/10.1007/JHEP01(2024)057} {\bibfield  {journal} {\bibinfo  {journal} {JHEP}\ }\textbf {\bibinfo {volume} {01}},\ \bibinfo {pages} {057}},\ \Eprint {https://arxiv.org/abs/2310.09152} {arXiv:2310.09152 [astro-ph.CO]} \BibitemShut {NoStop}%
\bibitem [{\citenamefont {Peccei}\ and\ \citenamefont {Quinn}(1977{\natexlab{a}})}]{Peccei:1977hh}%
  \BibitemOpen
  \bibfield  {author} {\bibinfo {author} {\bibfnamefont {R.~D.}\ \bibnamefont {Peccei}}\ and\ \bibinfo {author} {\bibfnamefont {H.~R.}\ \bibnamefont {Quinn}},\ }\bibfield  {title} {\bibinfo {title} {{CP Conservation in the Presence of Instantons}},\ }\href {https://doi.org/10.1103/PhysRevLett.38.1440} {\bibfield  {journal} {\bibinfo  {journal} {Phys. Rev. Lett.}\ }\textbf {\bibinfo {volume} {38}},\ \bibinfo {pages} {1440} (\bibinfo {year} {1977}{\natexlab{a}})}\BibitemShut {NoStop}%
\bibitem [{\citenamefont {Peccei}\ and\ \citenamefont {Quinn}(1977{\natexlab{b}})}]{Peccei:1977ur}%
  \BibitemOpen
  \bibfield  {author} {\bibinfo {author} {\bibfnamefont {R.~D.}\ \bibnamefont {Peccei}}\ and\ \bibinfo {author} {\bibfnamefont {H.~R.}\ \bibnamefont {Quinn}},\ }\bibfield  {title} {\bibinfo {title} {{Constraints Imposed by CP Conservation in the Presence of Instantons}},\ }\href {https://doi.org/10.1103/PhysRevD.16.1791} {\bibfield  {journal} {\bibinfo  {journal} {Phys. Rev. D}\ }\textbf {\bibinfo {volume} {16}},\ \bibinfo {pages} {1791} (\bibinfo {year} {1977}{\natexlab{b}})}\BibitemShut {NoStop}%
\bibitem [{\citenamefont {Weinberg}(1978)}]{Weinberg:1977ma}%
  \BibitemOpen
  \bibfield  {author} {\bibinfo {author} {\bibfnamefont {S.}~\bibnamefont {Weinberg}},\ }\bibfield  {title} {\bibinfo {title} {{A New Light Boson?}},\ }\href {https://doi.org/10.1103/PhysRevLett.40.223} {\bibfield  {journal} {\bibinfo  {journal} {Phys. Rev. Lett.}\ }\textbf {\bibinfo {volume} {40}},\ \bibinfo {pages} {223} (\bibinfo {year} {1978})}\BibitemShut {NoStop}%
\bibitem [{\citenamefont {Wilczek}(1978)}]{Wilczek:1977pj}%
  \BibitemOpen
  \bibfield  {author} {\bibinfo {author} {\bibfnamefont {F.}~\bibnamefont {Wilczek}},\ }\bibfield  {title} {\bibinfo {title} {{Problem of Strong $P$ and $T$ Invariance in the Presence of Instantons}},\ }\href {https://doi.org/10.1103/PhysRevLett.40.279} {\bibfield  {journal} {\bibinfo  {journal} {Phys. Rev. Lett.}\ }\textbf {\bibinfo {volume} {40}},\ \bibinfo {pages} {279} (\bibinfo {year} {1978})}\BibitemShut {NoStop}%
\bibitem [{\citenamefont {Aghanim}\ \emph {et~al.}(2020)\citenamefont {Aghanim} \emph {et~al.}}]{2020A&A...641A...6P}%
  \BibitemOpen
  \bibfield  {author} {\bibinfo {author} {\bibfnamefont {N.}~\bibnamefont {Aghanim}} \emph {et~al.} (\bibinfo {collaboration} {Planck}),\ }\bibfield  {title} {\bibinfo {title} {{Planck 2018 results. VI. Cosmological parameters}},\ }\href {https://doi.org/10.1051/0004-6361/201833910} {\bibfield  {journal} {\bibinfo  {journal} {Astron. Astrophys.}\ }\textbf {\bibinfo {volume} {641}},\ \bibinfo {pages} {A6} (\bibinfo {year} {2020})},\ \bibinfo {note} {[Erratum: Astron.Astrophys. 652, C4 (2021)]},\ \Eprint {https://arxiv.org/abs/1807.06209} {arXiv:1807.06209 [astro-ph.CO]} \BibitemShut {NoStop}%
\bibitem [{\citenamefont {Lagu\"e}\ \emph {et~al.}(2024)\citenamefont {Lagu\"e}, \citenamefont {Schwabe}, \citenamefont {Hlo\v{z}ek}, \citenamefont {Marsh},\ and\ \citenamefont {Rogers}}]{2024PhRvD.109d3507L}%
  \BibitemOpen
  \bibfield  {author} {\bibinfo {author} {\bibfnamefont {A.}~\bibnamefont {Lagu\"e}}, \bibinfo {author} {\bibfnamefont {B.}~\bibnamefont {Schwabe}}, \bibinfo {author} {\bibfnamefont {R.}~\bibnamefont {Hlo\v{z}ek}}, \bibinfo {author} {\bibfnamefont {D.~J.~E.}\ \bibnamefont {Marsh}},\ and\ \bibinfo {author} {\bibfnamefont {K.~K.}\ \bibnamefont {Rogers}},\ }\bibfield  {title} {\bibinfo {title} {{Cosmological simulations of mixed ultralight dark matter}},\ }\href {https://doi.org/10.1103/PhysRevD.109.043507} {\bibfield  {journal} {\bibinfo  {journal} {Phys. Rev. D}\ }\textbf {\bibinfo {volume} {109}},\ \bibinfo {pages} {043507} (\bibinfo {year} {2024})},\ \Eprint {https://arxiv.org/abs/2310.20000} {arXiv:2310.20000 [astro-ph.CO]} \BibitemShut {NoStop}%
\bibitem [{\citenamefont {Schwabe}\ \emph {et~al.}(2020)\citenamefont {Schwabe}, \citenamefont {Gosenca}, \citenamefont {Behrens}, \citenamefont {Niemeyer},\ and\ \citenamefont {Easther}}]{Schwabe:2020eac}%
  \BibitemOpen
  \bibfield  {author} {\bibinfo {author} {\bibfnamefont {B.}~\bibnamefont {Schwabe}}, \bibinfo {author} {\bibfnamefont {M.}~\bibnamefont {Gosenca}}, \bibinfo {author} {\bibfnamefont {C.}~\bibnamefont {Behrens}}, \bibinfo {author} {\bibfnamefont {J.~C.}\ \bibnamefont {Niemeyer}},\ and\ \bibinfo {author} {\bibfnamefont {R.}~\bibnamefont {Easther}},\ }\bibfield  {title} {\bibinfo {title} {{Simulating mixed fuzzy and cold dark matter}},\ }\href {https://doi.org/10.1103/PhysRevD.102.083518} {\bibfield  {journal} {\bibinfo  {journal} {Phys. Rev. D}\ }\textbf {\bibinfo {volume} {102}},\ \bibinfo {pages} {083518} (\bibinfo {year} {2020})},\ \Eprint {https://arxiv.org/abs/2007.08256} {arXiv:2007.08256 [astro-ph.CO]} \BibitemShut {NoStop}%
\bibitem [{\citenamefont {Hlozek}\ \emph {et~al.}(2015)\citenamefont {Hlozek}, \citenamefont {Grin}, \citenamefont {Marsh},\ and\ \citenamefont {Ferreira}}]{Hlozek:2014lca}%
  \BibitemOpen
  \bibfield  {author} {\bibinfo {author} {\bibfnamefont {R.}~\bibnamefont {Hlozek}}, \bibinfo {author} {\bibfnamefont {D.}~\bibnamefont {Grin}}, \bibinfo {author} {\bibfnamefont {D.~J.~E.}\ \bibnamefont {Marsh}},\ and\ \bibinfo {author} {\bibfnamefont {P.~G.}\ \bibnamefont {Ferreira}},\ }\bibfield  {title} {\bibinfo {title} {{A search for ultralight axions using precision cosmological data}},\ }\href {https://doi.org/10.1103/PhysRevD.91.103512} {\bibfield  {journal} {\bibinfo  {journal} {Phys. Rev. D}\ }\textbf {\bibinfo {volume} {91}},\ \bibinfo {pages} {103512} (\bibinfo {year} {2015})},\ \Eprint {https://arxiv.org/abs/1410.2896} {arXiv:1410.2896 [astro-ph.CO]} \BibitemShut {NoStop}%
\bibitem [{\citenamefont {Hlozek}\ \emph {et~al.}(2018)\citenamefont {Hlozek}, \citenamefont {Marsh},\ and\ \citenamefont {Grin}}]{Hlozek:2017zzf}%
  \BibitemOpen
  \bibfield  {author} {\bibinfo {author} {\bibfnamefont {R.}~\bibnamefont {Hlozek}}, \bibinfo {author} {\bibfnamefont {D.~J.~E.}\ \bibnamefont {Marsh}},\ and\ \bibinfo {author} {\bibfnamefont {D.}~\bibnamefont {Grin}},\ }\bibfield  {title} {\bibinfo {title} {{Using the Full Power of the Cosmic Microwave Background to Probe Axion Dark Matter}},\ }\href {https://doi.org/10.1093/mnras/sty271} {\bibfield  {journal} {\bibinfo  {journal} {Mon. Not. Roy. Astron. Soc.}\ }\textbf {\bibinfo {volume} {476}},\ \bibinfo {pages} {3063} (\bibinfo {year} {2018})},\ \Eprint {https://arxiv.org/abs/1708.05681} {arXiv:1708.05681 [astro-ph.CO]} \BibitemShut {NoStop}%
\bibitem [{\citenamefont {Lagu\"e}\ \emph {et~al.}(2022)\citenamefont {Lagu\"e}, \citenamefont {Bond}, \citenamefont {Hlo\v{z}ek}, \citenamefont {Rogers}, \citenamefont {Marsh},\ and\ \citenamefont {Grin}}]{Lague:2021frh}%
  \BibitemOpen
  \bibfield  {author} {\bibinfo {author} {\bibfnamefont {A.}~\bibnamefont {Lagu\"e}}, \bibinfo {author} {\bibfnamefont {J.~R.}\ \bibnamefont {Bond}}, \bibinfo {author} {\bibfnamefont {R.}~\bibnamefont {Hlo\v{z}ek}}, \bibinfo {author} {\bibfnamefont {K.~K.}\ \bibnamefont {Rogers}}, \bibinfo {author} {\bibfnamefont {D.~J.~E.}\ \bibnamefont {Marsh}},\ and\ \bibinfo {author} {\bibfnamefont {D.}~\bibnamefont {Grin}},\ }\bibfield  {title} {\bibinfo {title} {{Constraining ultralight axions with galaxy surveys}},\ }\href {https://doi.org/10.1088/1475-7516/2022/01/049} {\bibfield  {journal} {\bibinfo  {journal} {JCAP}\ }\textbf {\bibinfo {volume} {01}}\bibfield  {number} {\bibinfo  {number} { (01)},\ \bibinfo {pages} {049}},\ }\Eprint {https://arxiv.org/abs/2104.07802} {arXiv:2104.07802 [astro-ph.CO]} \BibitemShut {NoStop}%
\bibitem [{\citenamefont {Rogers}\ \emph {et~al.}(2023)\citenamefont {Rogers}, \citenamefont {Hlo\v{z}ek}, \citenamefont {Lagu\"e}, \citenamefont {Ivanov}, \citenamefont {Philcox}, \citenamefont {Cabass}, \citenamefont {Akitsu},\ and\ \citenamefont {Marsh}}]{Rogers:2023ezo}%
  \BibitemOpen
  \bibfield  {author} {\bibinfo {author} {\bibfnamefont {K.~K.}\ \bibnamefont {Rogers}}, \bibinfo {author} {\bibfnamefont {R.}~\bibnamefont {Hlo\v{z}ek}}, \bibinfo {author} {\bibfnamefont {A.}~\bibnamefont {Lagu\"e}}, \bibinfo {author} {\bibfnamefont {M.~M.}\ \bibnamefont {Ivanov}}, \bibinfo {author} {\bibfnamefont {O.~H.~E.}\ \bibnamefont {Philcox}}, \bibinfo {author} {\bibfnamefont {G.}~\bibnamefont {Cabass}}, \bibinfo {author} {\bibfnamefont {K.}~\bibnamefont {Akitsu}},\ and\ \bibinfo {author} {\bibfnamefont {D.~J.~E.}\ \bibnamefont {Marsh}},\ }\bibfield  {title} {\bibinfo {title} {{Ultra-light axions and the S $_{8}$ tension: joint constraints from the cosmic microwave background and galaxy clustering}},\ }\href {https://doi.org/10.1088/1475-7516/2023/06/023} {\bibfield  {journal} {\bibinfo  {journal} {JCAP}\ }\textbf {\bibinfo {volume} {06}},\ \bibinfo {pages} {023}},\ \Eprint {https://arxiv.org/abs/2301.08361} {arXiv:2301.08361 [astro-ph.CO]} \BibitemShut {NoStop}%
\bibitem [{\citenamefont {Centers}\ \emph {et~al.}(2021)\citenamefont {Centers} \emph {et~al.}}]{2019arXiv190513650C}%
  \BibitemOpen
  \bibfield  {author} {\bibinfo {author} {\bibfnamefont {G.~P.}\ \bibnamefont {Centers}} \emph {et~al.},\ }\bibfield  {title} {\bibinfo {title} {{Stochastic fluctuations of bosonic dark matter}},\ }\href {https://doi.org/10.1038/s41467-021-27632-7} {\bibfield  {journal} {\bibinfo  {journal} {Nature Commun.}\ }\textbf {\bibinfo {volume} {12}},\ \bibinfo {pages} {7321} (\bibinfo {year} {2021})},\ \Eprint {https://arxiv.org/abs/1905.13650} {arXiv:1905.13650 [astro-ph.CO]} \BibitemShut {NoStop}%
\bibitem [{\citenamefont {Nakatsuka}\ \emph {et~al.}(2023)\citenamefont {Nakatsuka}, \citenamefont {Morisaki}, \citenamefont {Fujita}, \citenamefont {Kume}, \citenamefont {Michimura}, \citenamefont {Nagano},\ and\ \citenamefont {Obata}}]{Nakatsuka:2022gaf}%
  \BibitemOpen
  \bibfield  {author} {\bibinfo {author} {\bibfnamefont {H.}~\bibnamefont {Nakatsuka}}, \bibinfo {author} {\bibfnamefont {S.}~\bibnamefont {Morisaki}}, \bibinfo {author} {\bibfnamefont {T.}~\bibnamefont {Fujita}}, \bibinfo {author} {\bibfnamefont {J.}~\bibnamefont {Kume}}, \bibinfo {author} {\bibfnamefont {Y.}~\bibnamefont {Michimura}}, \bibinfo {author} {\bibfnamefont {K.}~\bibnamefont {Nagano}},\ and\ \bibinfo {author} {\bibfnamefont {I.}~\bibnamefont {Obata}},\ }\bibfield  {title} {\bibinfo {title} {{Stochastic effects on observation of ultralight bosonic dark matter}},\ }\href {https://doi.org/10.1103/PhysRevD.108.092010} {\bibfield  {journal} {\bibinfo  {journal} {Phys. Rev. D}\ }\textbf {\bibinfo {volume} {108}},\ \bibinfo {pages} {092010} (\bibinfo {year} {2023})},\ \Eprint {https://arxiv.org/abs/2205.02960} {arXiv:2205.02960 [astro-ph.CO]} \BibitemShut {NoStop}%
\bibitem [{\citenamefont {Adachi}\ \emph {et~al.}(2023)\citenamefont {Adachi} \emph {et~al.}}]{2023PhRvD.108d3017A}%
  \BibitemOpen
  \bibfield  {author} {\bibinfo {author} {\bibfnamefont {S.}~\bibnamefont {Adachi}} \emph {et~al.} (\bibinfo {collaboration} {POLARBEAR}),\ }\bibfield  {title} {\bibinfo {title} {{Constraints on axionlike polarization oscillations in the cosmic microwave background with POLARBEAR}},\ }\href {https://doi.org/10.1103/PhysRevD.108.043017} {\bibfield  {journal} {\bibinfo  {journal} {Phys. Rev. D}\ }\textbf {\bibinfo {volume} {108}},\ \bibinfo {pages} {043017} (\bibinfo {year} {2023})},\ \Eprint {https://arxiv.org/abs/2303.08410} {arXiv:2303.08410 [astro-ph.CO]} \BibitemShut {NoStop}%
\bibitem [{\citenamefont {Iocco}\ \emph {et~al.}(2015)\citenamefont {Iocco}, \citenamefont {Pato},\ and\ \citenamefont {Bertone}}]{2015NatPh..11..245I}%
  \BibitemOpen
  \bibfield  {author} {\bibinfo {author} {\bibfnamefont {F.}~\bibnamefont {Iocco}}, \bibinfo {author} {\bibfnamefont {M.}~\bibnamefont {Pato}},\ and\ \bibinfo {author} {\bibfnamefont {G.}~\bibnamefont {Bertone}},\ }\bibfield  {title} {\bibinfo {title} {{Evidence for dark matter in the inner Milky Way}},\ }\href {https://doi.org/10.1038/nphys3237} {\bibfield  {journal} {\bibinfo  {journal} {Nature Phys.}\ }\textbf {\bibinfo {volume} {11}},\ \bibinfo {pages} {245} (\bibinfo {year} {2015})},\ \Eprint {https://arxiv.org/abs/1502.03821} {arXiv:1502.03821 [astro-ph.GA]} \BibitemShut {NoStop}%
\bibitem [{\citenamefont {Sivertsson}\ \emph {et~al.}(2018)\citenamefont {Sivertsson}, \citenamefont {Silverwood}, \citenamefont {Read}, \citenamefont {Bertone},\ and\ \citenamefont {Steger}}]{2018MNRAS.478.1677S}%
  \BibitemOpen
  \bibfield  {author} {\bibinfo {author} {\bibfnamefont {S.}~\bibnamefont {Sivertsson}}, \bibinfo {author} {\bibfnamefont {H.}~\bibnamefont {Silverwood}}, \bibinfo {author} {\bibfnamefont {J.~I.}\ \bibnamefont {Read}}, \bibinfo {author} {\bibfnamefont {G.}~\bibnamefont {Bertone}},\ and\ \bibinfo {author} {\bibfnamefont {P.}~\bibnamefont {Steger}},\ }\bibfield  {title} {\bibinfo {title} {{The localdark matter density from SDSS-SEGUE G-dwarfs}},\ }\href {https://doi.org/10.1093/mnras/sty977} {\bibfield  {journal} {\bibinfo  {journal} {Mon. Not. Roy. Astron. Soc.}\ }\textbf {\bibinfo {volume} {478}},\ \bibinfo {pages} {1677} (\bibinfo {year} {2018})},\ \Eprint {https://arxiv.org/abs/1708.07836} {arXiv:1708.07836 [astro-ph.GA]} \BibitemShut {NoStop}%
\bibitem [{\citenamefont {Anastassopoulos}\ \emph {et~al.}(2017)\citenamefont {Anastassopoulos} \emph {et~al.}}]{2017NatPh..13..584A}%
  \BibitemOpen
  \bibfield  {author} {\bibinfo {author} {\bibfnamefont {V.}~\bibnamefont {Anastassopoulos}} \emph {et~al.} (\bibinfo {collaboration} {CAST}),\ }\bibfield  {title} {\bibinfo {title} {{New CAST Limit on the Axion-Photon Interaction}},\ }\href {https://doi.org/10.1038/nphys4109} {\bibfield  {journal} {\bibinfo  {journal} {Nature Phys.}\ }\textbf {\bibinfo {volume} {13}},\ \bibinfo {pages} {584} (\bibinfo {year} {2017})},\ \Eprint {https://arxiv.org/abs/1705.02290} {arXiv:1705.02290 [hep-ex]} \BibitemShut {NoStop}%
\bibitem [{\citenamefont {Niemeyer}(2020)}]{NIEMEYER2020103787}%
  \BibitemOpen
  \bibfield  {author} {\bibinfo {author} {\bibfnamefont {J.~C.}\ \bibnamefont {Niemeyer}},\ }\bibfield  {title} {\bibinfo {title} {Small-scale structure of fuzzy and axion-like dark matter},\ }\href {https://doi.org/https://doi.org/10.1016/j.ppnp.2020.103787} {\bibfield  {journal} {\bibinfo  {journal} {Progress in Particle and Nuclear Physics}\ }\textbf {\bibinfo {volume} {113}},\ \bibinfo {pages} {103787} (\bibinfo {year} {2020})}\BibitemShut {NoStop}%
\bibitem [{\citenamefont {Ferreira}(2021)}]{2021A&ARv..29....7F}%
  \BibitemOpen
  \bibfield  {author} {\bibinfo {author} {\bibfnamefont {E.~G.~M.}\ \bibnamefont {Ferreira}},\ }\bibfield  {title} {\bibinfo {title} {{Ultra-light dark matter}},\ }\href {https://doi.org/10.1007/s00159-021-00135-6} {\bibfield  {journal} {\bibinfo  {journal} {Astron. Astrophys. Rev.}\ }\textbf {\bibinfo {volume} {29}},\ \bibinfo {pages} {7} (\bibinfo {year} {2021})},\ \Eprint {https://arxiv.org/abs/2005.03254} {arXiv:2005.03254 [astro-ph.CO]} \BibitemShut {NoStop}%
\bibitem [{\citenamefont {Hui}(2021)}]{Hui:2021tkt}%
  \BibitemOpen
  \bibfield  {author} {\bibinfo {author} {\bibfnamefont {L.}~\bibnamefont {Hui}},\ }\bibfield  {title} {\bibinfo {title} {{Wave Dark Matter}},\ }\href {https://doi.org/10.1146/annurev-astro-120920-010024} {\bibfield  {journal} {\bibinfo  {journal} {Ann. Rev. Astron. Astrophys.}\ }\textbf {\bibinfo {volume} {59}},\ \bibinfo {pages} {247} (\bibinfo {year} {2021})},\ \Eprint {https://arxiv.org/abs/2101.11735} {arXiv:2101.11735 [astro-ph.CO]} \BibitemShut {NoStop}%
\bibitem [{\citenamefont {Adams}\ \emph {et~al.}(2022)\citenamefont {Adams} \emph {et~al.}}]{Adams:2022pbo}%
  \BibitemOpen
  \bibfield  {author} {\bibinfo {author} {\bibfnamefont {C.~B.}\ \bibnamefont {Adams}} \emph {et~al.},\ }\bibfield  {title} {\bibinfo {title} {{Axion Dark Matter}},\ }in\ \href@noop {} {\emph {\bibinfo {booktitle} {{Snowmass 2021}}}}\ (\bibinfo {year} {2022})\ \Eprint {https://arxiv.org/abs/2203.14923} {arXiv:2203.14923 [hep-ex]} \BibitemShut {NoStop}%
\bibitem [{\citenamefont {Ir\v{s}i\v{c}}\ \emph {et~al.}(2017)\citenamefont {Ir\v{s}i\v{c}}, \citenamefont {Viel}, \citenamefont {Haehnelt}, \citenamefont {Bolton},\ and\ \citenamefont {Becker}}]{Irsic:2017yje}%
  \BibitemOpen
  \bibfield  {author} {\bibinfo {author} {\bibfnamefont {V.}~\bibnamefont {Ir\v{s}i\v{c}}}, \bibinfo {author} {\bibfnamefont {M.}~\bibnamefont {Viel}}, \bibinfo {author} {\bibfnamefont {M.~G.}\ \bibnamefont {Haehnelt}}, \bibinfo {author} {\bibfnamefont {J.~S.}\ \bibnamefont {Bolton}},\ and\ \bibinfo {author} {\bibfnamefont {G.~D.}\ \bibnamefont {Becker}},\ }\bibfield  {title} {\bibinfo {title} {{First constraints on fuzzy dark matter from Lyman-$\alpha$ forest data and hydrodynamical simulations}},\ }\href {https://doi.org/10.1103/PhysRevLett.119.031302} {\bibfield  {journal} {\bibinfo  {journal} {Phys. Rev. Lett.}\ }\textbf {\bibinfo {volume} {119}},\ \bibinfo {pages} {031302} (\bibinfo {year} {2017})},\ \Eprint {https://arxiv.org/abs/1703.04683} {arXiv:1703.04683 [astro-ph.CO]} \BibitemShut {NoStop}%
\bibitem [{\citenamefont {Kobayashi}\ \emph {et~al.}(2017)\citenamefont {Kobayashi}, \citenamefont {Murgia}, \citenamefont {De~Simone}, \citenamefont {Ir\v{s}i\v{c}},\ and\ \citenamefont {Viel}}]{Kobayashi:2017jcf}%
  \BibitemOpen
  \bibfield  {author} {\bibinfo {author} {\bibfnamefont {T.}~\bibnamefont {Kobayashi}}, \bibinfo {author} {\bibfnamefont {R.}~\bibnamefont {Murgia}}, \bibinfo {author} {\bibfnamefont {A.}~\bibnamefont {De~Simone}}, \bibinfo {author} {\bibfnamefont {V.}~\bibnamefont {Ir\v{s}i\v{c}}},\ and\ \bibinfo {author} {\bibfnamefont {M.}~\bibnamefont {Viel}},\ }\bibfield  {title} {\bibinfo {title} {{Lyman-$\alpha$ constraints on ultralight scalar dark matter: Implications for the early and late universe}},\ }\href {https://doi.org/10.1103/PhysRevD.96.123514} {\bibfield  {journal} {\bibinfo  {journal} {Phys. Rev. D}\ }\textbf {\bibinfo {volume} {96}},\ \bibinfo {pages} {123514} (\bibinfo {year} {2017})},\ \Eprint {https://arxiv.org/abs/1708.00015} {arXiv:1708.00015 [astro-ph.CO]} \BibitemShut {NoStop}%
\bibitem [{\citenamefont {Rogers}\ and\ \citenamefont {Peiris}(2021)}]{Rogers:2020ltq}%
  \BibitemOpen
  \bibfield  {author} {\bibinfo {author} {\bibfnamefont {K.~K.}\ \bibnamefont {Rogers}}\ and\ \bibinfo {author} {\bibfnamefont {H.~V.}\ \bibnamefont {Peiris}},\ }\bibfield  {title} {\bibinfo {title} {{Strong Bound on Canonical Ultralight Axion Dark Matter from the Lyman-Alpha Forest}},\ }\href {https://doi.org/10.1103/PhysRevLett.126.071302} {\bibfield  {journal} {\bibinfo  {journal} {Phys. Rev. Lett.}\ }\textbf {\bibinfo {volume} {126}},\ \bibinfo {pages} {071302} (\bibinfo {year} {2021})},\ \Eprint {https://arxiv.org/abs/2007.12705} {arXiv:2007.12705 [astro-ph.CO]} \BibitemShut {NoStop}%
\bibitem [{\citenamefont {L{\'o}pez}\ \emph {et~al.}(2016)\citenamefont {L{\'o}pez}, \citenamefont {D’Odorico}, \citenamefont {Ellison}, \citenamefont {Becker}, \citenamefont {Christensen}, \citenamefont {Cupani}, \citenamefont {Denney}, \citenamefont {P{\^a}ris}, \citenamefont {Worseck}, \citenamefont {Berg} \emph {et~al.}}]{2016A&A...594A..91L}%
  \BibitemOpen
  \bibfield  {author} {\bibinfo {author} {\bibfnamefont {S.}~\bibnamefont {L{\'o}pez}}, \bibinfo {author} {\bibfnamefont {V.}~\bibnamefont {D’Odorico}}, \bibinfo {author} {\bibfnamefont {S.}~\bibnamefont {Ellison}}, \bibinfo {author} {\bibfnamefont {G.}~\bibnamefont {Becker}}, \bibinfo {author} {\bibfnamefont {L.}~\bibnamefont {Christensen}}, \bibinfo {author} {\bibfnamefont {G.}~\bibnamefont {Cupani}}, \bibinfo {author} {\bibfnamefont {K.}~\bibnamefont {Denney}}, \bibinfo {author} {\bibfnamefont {I.}~\bibnamefont {P{\^a}ris}}, \bibinfo {author} {\bibfnamefont {G.}~\bibnamefont {Worseck}}, \bibinfo {author} {\bibfnamefont {T.}~\bibnamefont {Berg}}, \emph {et~al.},\ }\bibfield  {title} {\bibinfo {title} {Xq-100: A legacy survey of one hundred 3.5 {\ensuremath{\lesssim}} z {\ensuremath{\lesssim}} 4.5 quasars observed with vlt/x-shooter},\ }\href {https://doi.org/10.1051/0004-6361/201628161} {\bibfield  {journal} {\bibinfo  {journal} {Astronomy \& Astrophysics}\ }\textbf {\bibinfo {volume} {594}},\ \bibinfo
  {pages} {A91} (\bibinfo {year} {2016})},\ \Eprint {https://arxiv.org/abs/1607.08776} {arXiv:1607.08776 [astro-ph.GA]} \BibitemShut {NoStop}%
\bibitem [{\citenamefont {Winch}\ \emph {et~al.}(2024)\citenamefont {Winch}, \citenamefont {Rogers}, \citenamefont {Hlo{\v{z}}ek},\ and\ \citenamefont {Marsh}}]{2024arXiv240411071W}%
  \BibitemOpen
  \bibfield  {author} {\bibinfo {author} {\bibfnamefont {H.}~\bibnamefont {Winch}}, \bibinfo {author} {\bibfnamefont {K.~K.}\ \bibnamefont {Rogers}}, \bibinfo {author} {\bibfnamefont {R.}~\bibnamefont {Hlo{\v{z}}ek}},\ and\ \bibinfo {author} {\bibfnamefont {D.~J.}\ \bibnamefont {Marsh}},\ }\bibfield  {title} {\bibinfo {title} {High-redshift, small-scale tests of ultralight axion dark matter using hubble and webb galaxy uv luminosities},\ }\href@noop {} {\bibfield  {journal} {\bibinfo  {journal} {arXiv preprint arXiv:2404.11071}\ } (\bibinfo {year} {2024})}\BibitemShut {NoStop}%
\bibitem [{\citenamefont {Hayashi}\ \emph {et~al.}(2021)\citenamefont {Hayashi}, \citenamefont {Ferreira},\ and\ \citenamefont {Chan}}]{Hayashi:2021xxu}%
  \BibitemOpen
  \bibfield  {author} {\bibinfo {author} {\bibfnamefont {K.}~\bibnamefont {Hayashi}}, \bibinfo {author} {\bibfnamefont {E.~G.~M.}\ \bibnamefont {Ferreira}},\ and\ \bibinfo {author} {\bibfnamefont {H.~Y.~J.}\ \bibnamefont {Chan}},\ }\bibfield  {title} {\bibinfo {title} {{Narrowing the Mass Range of Fuzzy Dark Matter with Ultrafaint Dwarfs}},\ }\href {https://doi.org/10.3847/2041-8213/abf501} {\bibfield  {journal} {\bibinfo  {journal} {Astrophys. J. Lett.}\ }\textbf {\bibinfo {volume} {912}},\ \bibinfo {pages} {L3} (\bibinfo {year} {2021})},\ \Eprint {https://arxiv.org/abs/2102.05300} {arXiv:2102.05300 [astro-ph.CO]} \BibitemShut {NoStop}%
\bibitem [{\citenamefont {Zimmermann}\ \emph {et~al.}(2024)\citenamefont {Zimmermann}, \citenamefont {Alvey}, \citenamefont {Marsh}, \citenamefont {Fairbairn},\ and\ \citenamefont {Read}}]{Zimmermann:2024xvd}%
  \BibitemOpen
  \bibfield  {author} {\bibinfo {author} {\bibfnamefont {T.}~\bibnamefont {Zimmermann}}, \bibinfo {author} {\bibfnamefont {J.}~\bibnamefont {Alvey}}, \bibinfo {author} {\bibfnamefont {D.~J.}\ \bibnamefont {Marsh}}, \bibinfo {author} {\bibfnamefont {M.}~\bibnamefont {Fairbairn}},\ and\ \bibinfo {author} {\bibfnamefont {J.~I.}\ \bibnamefont {Read}},\ }\bibfield  {title} {\bibinfo {title} {Dwarf galaxies imply dark matter is heavier than $\mathbf{2.2 \times 10^{-21}} \, \mathbf{eV}$},\ }\href@noop {} {\bibfield  {journal} {\bibinfo  {journal} {arXiv preprint arXiv:2405.20374}\ } (\bibinfo {year} {2024})}\BibitemShut {NoStop}%
\bibitem [{\citenamefont {Chan}\ \emph {et~al.}(2022)\citenamefont {Chan}, \citenamefont {Ferreira}, \citenamefont {May}, \citenamefont {Hayashi},\ and\ \citenamefont {Chiba}}]{Chan:2021bja}%
  \BibitemOpen
  \bibfield  {author} {\bibinfo {author} {\bibfnamefont {H.~Y.~J.}\ \bibnamefont {Chan}}, \bibinfo {author} {\bibfnamefont {E.~G.~M.}\ \bibnamefont {Ferreira}}, \bibinfo {author} {\bibfnamefont {S.}~\bibnamefont {May}}, \bibinfo {author} {\bibfnamefont {K.}~\bibnamefont {Hayashi}},\ and\ \bibinfo {author} {\bibfnamefont {M.}~\bibnamefont {Chiba}},\ }\bibfield  {title} {\bibinfo {title} {{The diversity of core\textendash{}halo structure in the fuzzy dark matter model}},\ }\href {https://doi.org/10.1093/mnras/stac063} {\bibfield  {journal} {\bibinfo  {journal} {Mon. Not. Roy. Astron. Soc.}\ }\textbf {\bibinfo {volume} {511}},\ \bibinfo {pages} {943} (\bibinfo {year} {2022})},\ \Eprint {https://arxiv.org/abs/2110.11882} {arXiv:2110.11882 [astro-ph.CO]} \BibitemShut {NoStop}%
\bibitem [{\citenamefont {Powell}\ \emph {et~al.}(2023)\citenamefont {Powell}, \citenamefont {Vegetti}, \citenamefont {McKean}, \citenamefont {White}, \citenamefont {Ferreira}, \citenamefont {May},\ and\ \citenamefont {Spingola}}]{Powell:2023jns}%
  \BibitemOpen
  \bibfield  {author} {\bibinfo {author} {\bibfnamefont {D.~M.}\ \bibnamefont {Powell}}, \bibinfo {author} {\bibfnamefont {S.}~\bibnamefont {Vegetti}}, \bibinfo {author} {\bibfnamefont {J.~P.}\ \bibnamefont {McKean}}, \bibinfo {author} {\bibfnamefont {S.~D.~M.}\ \bibnamefont {White}}, \bibinfo {author} {\bibfnamefont {E.~G.~M.}\ \bibnamefont {Ferreira}}, \bibinfo {author} {\bibfnamefont {S.}~\bibnamefont {May}},\ and\ \bibinfo {author} {\bibfnamefont {C.}~\bibnamefont {Spingola}},\ }\bibfield  {title} {\bibinfo {title} {{A lensed radio jet at milli-arcsecond resolution \textendash{} II. Constraints on fuzzy dark matter from an extended gravitational arc}},\ }\href {https://doi.org/10.1093/mnrasl/slad074} {\bibfield  {journal} {\bibinfo  {journal} {Mon. Not. Roy. Astron. Soc.}\ }\textbf {\bibinfo {volume} {524}},\ \bibinfo {pages} {L84} (\bibinfo {year} {2023})},\ \Eprint {https://arxiv.org/abs/2302.10941} {arXiv:2302.10941 [astro-ph.CO]} \BibitemShut {NoStop}%
\bibitem [{\citenamefont {Laroche}\ \emph {et~al.}(2022)\citenamefont {Laroche}, \citenamefont {Gilman}, \citenamefont {Li}, \citenamefont {Bovy},\ and\ \citenamefont {Du}}]{Laroche:2022pjm}%
  \BibitemOpen
  \bibfield  {author} {\bibinfo {author} {\bibfnamefont {A.}~\bibnamefont {Laroche}}, \bibinfo {author} {\bibfnamefont {D.}~\bibnamefont {Gilman}}, \bibinfo {author} {\bibfnamefont {X.}~\bibnamefont {Li}}, \bibinfo {author} {\bibfnamefont {J.}~\bibnamefont {Bovy}},\ and\ \bibinfo {author} {\bibfnamefont {X.}~\bibnamefont {Du}},\ }\bibfield  {title} {\bibinfo {title} {Quantum fluctuations masquerade as haloes: bounds on ultra-light dark matter from quadruply imaged quasars},\ }\href@noop {} {\bibfield  {journal} {\bibinfo  {journal} {Monthly Notices of the Royal Astronomical Society}\ }\textbf {\bibinfo {volume} {517}},\ \bibinfo {pages} {1867} (\bibinfo {year} {2022})}\BibitemShut {NoStop}%
\bibitem [{\citenamefont {Amruth}\ \emph {et~al.}(2023)\citenamefont {Amruth} \emph {et~al.}}]{Amruth:2023xqj}%
  \BibitemOpen
  \bibfield  {author} {\bibinfo {author} {\bibfnamefont {A.}~\bibnamefont {Amruth}} \emph {et~al.},\ }\bibfield  {title} {\bibinfo {title} {{Einstein rings modulated by wavelike dark matter from anomalies in gravitationally lensed images}},\ }\href {https://doi.org/10.1038/s41550-023-01943-9} {\bibfield  {journal} {\bibinfo  {journal} {Nature Astron.}\ }\textbf {\bibinfo {volume} {7}},\ \bibinfo {pages} {736} (\bibinfo {year} {2023})},\ \Eprint {https://arxiv.org/abs/2304.09895} {arXiv:2304.09895 [astro-ph.CO]} \BibitemShut {NoStop}%
\bibitem [{\citenamefont {Dalal}\ and\ \citenamefont {Kravtsov}(2022)}]{Dalal:2022rmp}%
  \BibitemOpen
  \bibfield  {author} {\bibinfo {author} {\bibfnamefont {N.}~\bibnamefont {Dalal}}\ and\ \bibinfo {author} {\bibfnamefont {A.}~\bibnamefont {Kravtsov}},\ }\bibfield  {title} {\bibinfo {title} {{Excluding fuzzy dark matter with sizes and stellar kinematics of ultrafaint dwarf galaxies}},\ }\href {https://doi.org/10.1103/PhysRevD.106.063517} {\bibfield  {journal} {\bibinfo  {journal} {Phys. Rev. D}\ }\textbf {\bibinfo {volume} {106}},\ \bibinfo {pages} {063517} (\bibinfo {year} {2022})},\ \Eprint {https://arxiv.org/abs/2203.05750} {arXiv:2203.05750 [astro-ph.CO]} \BibitemShut {NoStop}%
\bibitem [{\citenamefont {Ferguson}\ \emph {et~al.}(2022)\citenamefont {Ferguson} \emph {et~al.}}]{2022PhRvD.106d2011F}%
  \BibitemOpen
  \bibfield  {author} {\bibinfo {author} {\bibfnamefont {K.~R.}\ \bibnamefont {Ferguson}} \emph {et~al.} (\bibinfo {collaboration} {SPT-3G}),\ }\bibfield  {title} {\bibinfo {title} {{Searching for axionlike time-dependent cosmic birefringence with data from SPT-3G}},\ }\href {https://doi.org/10.1103/PhysRevD.106.042011} {\bibfield  {journal} {\bibinfo  {journal} {Phys. Rev. D}\ }\textbf {\bibinfo {volume} {106}},\ \bibinfo {pages} {042011} (\bibinfo {year} {2022})},\ \Eprint {https://arxiv.org/abs/2203.16567} {arXiv:2203.16567 [astro-ph.CO]} \BibitemShut {NoStop}%
\bibitem [{\citenamefont {Ade}\ \emph {et~al.}(2022)\citenamefont {Ade} \emph {et~al.}}]{2022PhRvD.105b2006A}%
  \BibitemOpen
  \bibfield  {author} {\bibinfo {author} {\bibfnamefont {P.~A.~R.}\ \bibnamefont {Ade}} \emph {et~al.} (\bibinfo {collaboration} {BICEP/Keck}),\ }\bibfield  {title} {\bibinfo {title} {{BICEP/Keck XIV: Improved constraints on axionlike polarization oscillations in the cosmic microwave background}},\ }\href {https://doi.org/10.1103/PhysRevD.105.022006} {\bibfield  {journal} {\bibinfo  {journal} {Phys. Rev. D}\ }\textbf {\bibinfo {volume} {105}},\ \bibinfo {pages} {022006} (\bibinfo {year} {2022})},\ \Eprint {https://arxiv.org/abs/2108.03316} {arXiv:2108.03316 [astro-ph.CO]} \BibitemShut {NoStop}%
\bibitem [{\citenamefont {Lewis}\ \emph {et~al.}(2000)\citenamefont {Lewis}, \citenamefont {Challinor},\ and\ \citenamefont {Lasenby}}]{Lewis:1999bs}%
  \BibitemOpen
  \bibfield  {author} {\bibinfo {author} {\bibfnamefont {A.}~\bibnamefont {Lewis}}, \bibinfo {author} {\bibfnamefont {A.}~\bibnamefont {Challinor}},\ and\ \bibinfo {author} {\bibfnamefont {A.}~\bibnamefont {Lasenby}},\ }\bibfield  {title} {\bibinfo {title} {Efficient computation of cmb anisotropies in closed frw models},\ }\href@noop {} {\bibfield  {journal} {\bibinfo  {journal} {\apj}\ }\textbf {\bibinfo {volume} {538}},\ \bibinfo {pages} {473} (\bibinfo {year} {2000})},\ \Eprint {https://arxiv.org/abs/astro-ph/9911177} {astro-ph/9911177} \BibitemShut {NoStop}%
\bibitem [{\citenamefont {Diego-Palazuelos}\ \emph {et~al.}(2024)\citenamefont {Diego-Palazuelos}, \citenamefont {de~Belsunce}, \citenamefont {Gratton},\ and\ \citenamefont {Sherwin}}]{Diego-Palazuelos:2024lym}%
  \BibitemOpen
  \bibfield  {author} {\bibinfo {author} {\bibfnamefont {P.}~\bibnamefont {Diego-Palazuelos}}, \bibinfo {author} {\bibfnamefont {R.}~\bibnamefont {de~Belsunce}}, \bibinfo {author} {\bibfnamefont {S.}~\bibnamefont {Gratton}},\ and\ \bibinfo {author} {\bibfnamefont {B.~D.}\ \bibnamefont {Sherwin}},\ }\bibfield  {title} {\bibinfo {title} {{Axion field tomography: cosmic birefringence from the epochs of recombination and reionization}},\ }\href {https://doi.org/10.22323/1.454.0047} {\bibfield  {journal} {\bibinfo  {journal} {PoS}\ }\textbf {\bibinfo {volume} {COSMICWISPers}},\ \bibinfo {pages} {047} (\bibinfo {year} {2024})}\BibitemShut {NoStop}%
\end{thebibliography}%

\end{document}